%% file: MSSM_Paper.tex
\documentclass[11pt,a4paper]{article} %for JHEP

\usepackage{atlasphysics}
\usepackage{jheppub}
\usepackage{subfigure}
\usepackage{units}
\usepackage{amsmath,graphicx}
\usepackage{booktabs}
\usepackage{authblk}
\usepackage{lineno}
\usepackage{multirow,xspace}
\usepackage{wasysym}
\usepackage{rotating}
\usepackage[normalem]{ulem}

\newcommand{\err}[2]{\ensuremath{^{+#1}_{-#2}}}

\graphicspath{{./figures/}}
\def\TeV{\ifmmode {\mathrm{Te\kern -0.1em V}}\else
                   \textrm{Te\kern -0.1em V}\fi}%
\def\GeV{\ifmmode {\mathrm{Ge\kern -0.1em V}}\else
                   \textrm{Ge\kern -0.1em V}\fi}%
\def\MeV{\ifmmode {\mathrm{Me\kern -0.1em V}}\else
                   \textrm{Me\kern -0.1em V}\fi}%
\def\keV{\ifmmode {\mathrm{ke\kern -0.1em V}}\else
                   \textrm{ke\kern -0.1em V}\fi}%
\def\eV{\ifmmode  {\mathrm{e\kern -0.1em V}}\else
                   \textrm{e\kern -0.1em V}\fi}%

\def\TeVc{\ifmmode {\mathrm{Te\kern -0.1em V}/c}\else
                   {\textrm{Te\kern -0.1em V}/$c$}\fi}%
\def\GeVc{\ifmmode {\mathrm{Ge\kern -0.1em V}/c}\else
                   {\textrm{Ge\kern -0.1em V}/$c$}\fi}%
\def\MeVc{\ifmmode {\mathrm{Me\kern -0.1em V}/c}\else
                   {\textrm{Me\kern -0.1em V}/$c$}\fi}%
\def\keVc{\ifmmode {\mathrm{ke\kern -0.1em V}/c}\else
                   {\textrm{ke\kern -0.1em V}/$c$}\fi}%
\def\eVc{\ifmmode  {\mathrm{e\kern -0.1em V}/c}\else
                   {\textrm{e\kern -0.1em V}/$c$}\fi}%

\def\TeVcc{\ifmmode {\mathrm{Te\kern -0.1em V}/c^2}\else
                   {\textrm{Te\kern -0.1em V}/$c^2$}\fi}%
\def\GeVcc{\ifmmode {\mathrm{Ge\kern -0.1em V}/c^2}\else
                   {\textrm{Ge\kern -0.1em V}/$c^2$}\fi}%
\def\MeVcc{\ifmmode {\mathrm{Me\kern -0.1em V}/c^2}\else
                   {\textrm{Me\kern -0.1em V}/$c^2$}\fi}%
\def\keVcc{\ifmmode {\mathrm{ke\kern -0.1em V}/c^2}\else
                   {\textrm{ke\kern -0.1em V}/$c^2$}\fi}%
\def\eVcc{\ifmmode  {\mathrm{e\kern -0.1em V}/c^2}\else
                   {\textrm{e\kern -0.1em V}/$c^2$}\fi}%

\def\GEVc{\GeV}
\def\GEVcc{\GeV}

\newcommand{\mT}{\ensuremath{m_{\rm T}}}

\def\MA{\ensuremath{m_{A}}}
\def\tanb{\ensuremath{\tan \beta}}
\newcommand{\tb}{\ensuremath{\tan\beta}}
\def\MET{\ensuremath{E_{\mathrm{T}}^{\mathrm{miss}}}}
\def\met{\ensuremath{E_{\mathrm T}^{\mathrm{miss}}}}

\def\pt{\ensuremath{p_{\mathrm{T}}}}

\newcommand{\HTlep}{\ensuremath{\met + p_{\text{T},e} + p_{\text{T},\mu}}}
\newcommand{\mmcmass}{\ensuremath{m^\text{MMC}_{\tau\tau}}}

\newcommand{\sumcos}{\ensuremath{\sum_{\ell=e,\mu}\cos\Delta\phi_{\met,\ell}}}
\newcommand{\deltaPhi}{\ensuremath{\Delta \phi_{e\mu}}}
\def\atau{\ensuremath{\tau}}
\def\tautau{\ensuremath{\tau^{+} \tau^{-}}}

\def\tauleptons{\ensuremath{\tau\,\mathrm{leptons}}}
\def\mumu{\ensuremath{\mu^{+} \mu^{-}}}
\def\ee{\ensuremath{e^{+} e^{-}}}

\def\thad{\ensuremath{\tau_{\mathrm{ had}}}}
\def\tlep{\ensuremath{\tau_{\mathrm{ lep}}}}
\def\telec{\ensuremath{\tau_{e}}}
\def\tmuon{\ensuremath{\tau_{\mu}}}

\def\bjet{$b$-jet}
\def\bjets{$b$-jets}

\def\bquark{$b$-quark}
\def\bquarks{$b$-quarks}

\def\Higgses{\ensuremath{h/A/H}}
\def\Htotautautolh{\ensuremath{\Higgses \to \tlep\thad\ }}
\def\Htotautautohh{\ensuremath{\Higgses \to \thad\thad\ }}
\def\Htotautautohhnospace{\ensuremath{\Higgses \to \thad\thad}}
\def\Htomumu{\ensuremath{\Higgses\to\mumu}}
\newcommand{\Amumu}{\Htomumu}
\def\Htotautau{\ensuremath{\Higgses\to\tautau}}
\def\Htotautauemu{\ensuremath{\Higgses \to \telec\tmuon\ }}

\def\Htotautau{\ensuremath{ \Higgses \to \tautau }}

\def\Wjets{\ensuremath{W+\mathrm{jets}}}
\def\Wmunu{\ensuremath{W(\to\mu\nu){\rm+jets}}}
\def\Wtomunu{\Wmunu}
\def\Wtolnu{\ensuremath{W\to\ell\nu}}

\def\Wtaunu{\ensuremath{W(\to\tau\nu){\rm+jets}}}
\def\Wtotaunu{\Wtaunu}

\def\WW{\ensuremath{W^+W^-}}

\def\Zgstar{\ensuremath{Z/\gamma^{*}}}
\def\Zjets{\ensuremath{Z/\gamma^{*} +\mathrm{jets}}}

\def\Ztotautau{\ensuremath{Z/\gamma^{*} \to \tautau}}
\def\Ztoll{\ensuremath{Z/\gamma^{*} \to \ell^+\ell^-}}
\def\Ztoee{\ensuremath{Z/\gamma^{*} \to \ee}}
\def\Ztomumu{\ensuremath{Z/\gamma^{*} \to \mumu}}

\def\ttbar{\ensuremath{t\bar{t}}}
\def\bbbar{\ensuremath{b\bar{b}}}
\def\TTBAR{\ensuremath{t\bar{t}}}

\def\mhmax{\ensuremath{m_h^{\mathrm{max}}}}
\def\currentlumi{$4.7\,\ifb$}
\def\currentlumirange{\unit[4.7]{\ifb} to \unit[4.8]{\ifb}}

\newcommand{\figureword}{figure}
\newcommand{\Figureword}{Figure} %beginning of sentence

\newcommand{\sectionword}{section}
\newcommand{\tableword}{table}
\newcommand{\Tableword}{Table} %beginning of sentence
\newcommand{\referenceword}{reference}
\newcommand{\refcite}[1]{\referenceword~\cite{#1}}
\newcommand{\refscite}[1]{\referenceword s~\cite{#1}}
\newcommand{\secref}[1]{\sectionword~\ref{#1}}
\newcommand{\tabref}[1]{\tableword~\ref{#1}}
\newcommand{\Tabref}[1]{\Tableword~\ref{#1}} %at beginning of sentence
\newcommand{\figref}[1]{\figureword~\ref{#1}}
\newcommand{\Figref}[1]{\Figureword~\ref{#1}} %at beginning of sentence

\newcommand{\percent}[1]{#1\%}

\mathchardef\mhyphen="2D

\usepackage{preprintcover}  % load package                                                                                            
\PreprintCoverPaperTitle{
\begin{boldmath}
\textbf{ 
Search for the neutral Higgs bosons of the Minimal Supersymmetric Standard Model in 
$\boldsymbol{pp}$ collisions at $\boldsymbol{\sqrt{s}=7\,\TeV}$ with the ATLAS detector
}
\end{boldmath}
}  % paper title       

\PreprintIdNumber{CERN-PH-EP-2012-323}  % CERN preprint number 
\PreprintCoverAbstract{
A search for neutral Higgs bosons of the Minimal Supersymmetric Standard Model (MSSM) is reported. The analysis is based on a sample of proton--proton collisions at a centre-of-mass energy of
\unit[7]{\TeV} recorded with the ATLAS detector at the Large Hadron Collider. The data were recorded in 2011 and correspond to an integrated luminosity of \currentlumirange. Higgs boson decays into 
oppositely-charged muon or $\tau$ lepton pairs are considered for final states requiring either the presence or absence of \bjets.  
No statistically significant excess over the
expected background is observed and exclusion limits at the \percent{95} confidence level are derived. The exclusion limits are for
the production cross-section of a generic neutral Higgs boson, $\phi$, as a function of the Higgs boson
mass and for \Higgses{} production in the MSSM as a function of the parameters $\MA$ and
$\tan\beta$ in the \mhmax{} scenario for $\MA$ in the range of \unit[90]{\GEVcc} to \unit[500]{\GEVcc}.
}
\PreprintJournalName{JHEP}

\title{\textbf{
Search for the neutral Higgs bosons of the Minimal Supersymmetric Standard Model in 
$\boldsymbol{pp}$ collisions at $\boldsymbol{\sqrt{s}=7\,\TeV}$ with the ATLAS detector
}}

\author{The ATLAS Collaboration}

\abstract{
A search for neutral Higgs bosons of the Minimal Supersymmetric Standard Model (MSSM) is reported. The analysis is based on a sample of proton--proton collisions at a centre-of-mass energy of
\unit[7]{\TeV} recorded with the ATLAS detector at the Large Hadron Collider. The data were recorded in 2011 and correspond to an integrated luminosity of \currentlumirange. Higgs boson decays into 
oppositely-charged muon or $\tau$ lepton pairs are considered for final states requiring either the presence or absence of \bjets.  
No statistically significant excess over the
expected background is observed and exclusion limits at the \percent{95} confidence level are derived. The exclusion limits are for
the production cross-section of a generic neutral Higgs boson, $\phi$, as a function of the Higgs boson
mass and for \Higgses{} production in the MSSM as a function of the parameters $\MA$ and
$\tan\beta$ in the \mhmax{} scenario for $\MA$ in the range of \unit[90]{\GEVcc} to \unit[500]{\GEVcc}.
}

\begin{document}
\maketitle
\flushbottom 

% \linenumbers

\input{introduction}

\input{samples}

\input{objects}

\input{mumu_channel}

\input{tautau_channel}

\input{systematics}

\input{statistics}

\input{results}

\input{conclusions}

\input{acknowledgements}

\bibliographystyle{JHEP-2}
%\bibliography{MSSM_Paper}
\providecommand{\href}[2]{#2}\begingroup\raggedright\endgroup

\clearpage
\input{atlas_authlist}

\end{document}
\bye

%% file: introduction.tex
\section{Introduction} \label{sec:intro}

Discovering the mechanism responsible for electroweak symmetry breaking
is one of the major goals of the physics programme at the Large Hadron Collider~(LHC)~\cite{LHC}.
In the Standard Model this mechanism requires the existence of a single scalar particle, 
the Higgs boson~\cite{ENGLERT,HIGGS,HIGGS2,HIGGS3,Guralnik:1964eu}.
The recent discovery of a particle compatible with the 
Higgs boson at the LHC~\cite{ATLASLimit,CMSLimit} provides further evidence in support of this simple picture. 
Even if this recently discovered particle is shown to have properties very close to the 
Standard Model Higgs boson, there are still a number of problems that are not addressed. 
For instance, quantum corrections to the mass of the Higgs boson contain quadratic divergences.
This problem can be solved by introducing supersymmetry, a 
symmetry between fermions and bosons, by which the divergent corrections to the Higgs 
boson mass are cancelled.

In the Minimal Supersymmetric Standard Model (MSSM)~\cite{MSSM1, MSSM2},
two Higgs doublets are necessary, coupling separately to up-type and down-type fermions.
This results in five physical Higgs bosons, two of which are neutral and CP-even ($h$, $H$)\footnote{By convention the lighter CP-even Higgs boson is called $h$, the heavier CP-even Higgs boson is called $H$.}, 
one of which is neutral and CP-odd ($A$), and two of which are charged ($H^\pm$). 
At tree level their properties can be described in terms of two parameters,
typically chosen to be the mass of the CP-odd Higgs boson, $\MA$, and the ratio of
the vacuum expectation values of the two Higgs doublets, $\tan\beta$.  
In the MSSM, the Higgs boson couplings to  $\tau$ leptons and \bquarks{}
are strongly enhanced for a large part of the parameter space.
This is especially true for large values of $\tan\beta$, in which case the
decay of a Higgs boson to a pair of $\tau$ leptons or \bquarks{} and its production
in association with \bquarks{} play a much more important role than in the Standard Model.

The results presented in this paper are interpreted
in the context of the \mhmax\ benchmark scenario~\cite{MSSMmhmax}. 
In the $\mhmax$ scenario the parameters of the model are chosen such that the
mass of the lightest CP-even Higgs boson, $h$, is maximised for a given point in the $\MA$--$\tan\beta$ plane, under certain
assumptions. This guarantees conservative exclusion bounds from the LEP experiments. 
The sign of the Higgs sector bilinear coupling, $\mu$, is generally not constrained, 
but for the analysis presented in this paper $\mu>0$ is chosen as this is favoured by the measurements of the anomalous magnetic dipole moment of the muon 
\cite{Jegerlehner20091}.

The masses of the Higgs bosons in this scenario are such that for large values of $\tan\beta$ two of the three neutral
Higgs bosons are closely degenerate in mass: for $\MA\lesssim \unit[130]{\GEVcc}$, $m_{h} \simeq \MA$
and $m_{H}\simeq \unit[130]{\GEVcc}$, whereas for $\MA\gtrsim \unit[130]{\GEVcc}$, $m_{H} \simeq \MA$ and $m_{h}\simeq \unit[130]{\GEVcc}$.

The most common MSSM neutral Higgs boson production mechanisms at a hadron collider are the \bquark\ associated production and gluon-fusion processes, 
the latter process proceeding primarily through a \bquark\ loop for intermediate and high $\tan\beta$. 
Both processes have cross-sections that increase with $\tan\beta$, with the $b$-associated production process becoming dominant at high $\tan\beta$ values. 
The most common decay modes at high $\tan\beta$ are to a pair of \bquarks\ or \atau\ leptons, with branching ratios close to \percent{90} and \percent{10}, respectively, across the mass range considered.
The direct decay into two muons occurs rarely, with a branching ratio around \percent{0.04}, but offers a clean signature.

Previous searches for neutral MSSM Higgs bosons have been performed at LEP~\cite{LEPLimits},
the Tevatron~\cite{TevatronLimits1,TevatronLimits2,TevatronLimits3,TevatronHbb,CDFHbb,D0Hbb} 
and the LHC~\cite{ATLASLimit,CMSLimit}.  
The recently observed Higgs-boson-like particle at the LHC~\cite{ATLASHiggsObservationJuly2012, CMSHiggsObservationJuly2012} is consistent
with both the Standard Model and the lightest
CP-even MSSM Higgs boson~\cite{theo125,theo125_v2}.
In this paper a search for neutral MSSM Higgs bosons using \currentlumirange{} of proton--proton
collision data collected with the ATLAS detector \cite{ATLASDetector} in 2011 at the centre-of-mass energy of \unit[7]{\TeV} is presented. The $\mumu$ and $\tautau$ decay modes are considered,
with the latter divided into separate search channels according to the subsequent $\tau$ lepton decay modes. 
Events from each channel are further classified according to the presence or the
absence of an identified \bjet.

%% file: samples.tex
\section{The ATLAS detector} 

The ATLAS experiment at the LHC is a multi-purpose particle detector with a forward-backward
symmetric cylindrical geometry and nearly $4\pi$ coverage in solid
angle~\cite{ATLASDetector}. It consists of an inner tracking detector surrounded by a thin
superconducting solenoid providing a \unit[2]{T} axial magnetic field, electromagnetic and
hadronic calorimeters, and a muon spectrometer.
The inner tracking detector
covers the pseudorapidity
range $|\eta| < 2.5$\footnote{
  ATLAS uses a right-handed coordinate system with
  its origin at the nominal interaction point (IP) in the centre of the detector
  and the $z$-axis along the beam pipe. The $x$-axis points from the IP to the centre of
  the LHC ring, and the $y$-axis points upwards. Cylindrical coordinates $(r,\phi)$ are used
  in the transverse plane, $\phi$ being the azimuthal angle around the beam pipe.
  The pseudorapidity is defined in terms of the polar angle $\theta$ as
  $\eta=-\ln\tan(\theta/2)$.}.
 It consists of silicon pixel, semi-conductor micro-strip, and transition radiation tracking detectors. Lead/liquid-argon (LAr) sampling calorimeters provide
electromagnetic (EM) energy measurements with high granularity. A hadronic (iron/scintillator-tile)
calorimeter covers the central pseudorapidity range ($|\eta| < 1.7$).
The end-cap and forward regions are instrumented with LAr calorimeters for both EM and
hadronic energy measurements up to $|\eta| = 4.9$. The muon spectrometer surrounds the calorimeters
and incorporates three large air-core toroid superconducting magnets with bending power between  \unit[2.0]{Tm} and \unit[7.5]{Tm},
a system of precision tracking chambers and fast detectors for triggering.
A three-level trigger system is used to select events. The first-level trigger is implemented
in hardware and uses a subset of the detector information to reduce the rate to
at most \unit[75]{kHz}. This is followed by two software-based trigger levels that together
reduce the event rate to approximately \unit[300]{Hz}. The trigger requirements were adjusted to
changing data-taking conditions during 2011.

\section{Data and Monte Carlo simulation samples} 
\label{sec:samples}
The data used in this search were recorded by the ATLAS experiment during the 2011 LHC run with proton--proton
collisions at a centre-of-mass energy of \unit[7]{\TeV}. They correspond to an integrated luminosity of
$\unit[4.7]{fb^{-1}}$ (\tautau{} channels) or $\unit[4.8]{fb^{-1}}$ (\mumu{}
channel) after imposing the
data quality selection criteria to require that all relevant detector sub-systems used in these analyses
were operational.

\paragraph{Higgs boson production:}
The Higgs boson production mechanisms considered are gluon-fusion and $b$-associated production.
The cross-sections for the first process have been calculated using
{\sc HIGLU}~\cite{HIGLU} 
and ggh@nnlo~\cite{ggh_at_nnlo}.
For $b$-associated production, 
a matching scheme described in \refcite{SantanderMatching} is used to combine  
four-flavour~\cite{Dittmaier:2003ej,Dawson:2004a} calculations and the 
five-flavour bbh@nlo~\cite{Harlander:2003ai} calculation. 
The masses, couplings and branching ratios
of the Higgs bosons are computed with FeynHiggs~\cite{feynhiggs}. 
Details of the calculations and the associated uncertainties due to the choice of the value of the strong coupling constant, the parton distribution 
function and the factorisation and renormalisation scales can be found in~\refcite{LHCHiggsCrossSectionWorkingGroup:2011ti}.
Gluon-fusion production is simulated with 
POWHEG~\cite{POWHEG}, while \bquark\ associated production is simulated with 
SHERPA~\cite{SHERPA}.

The \Htomumu{} and \Htotautau{} modes are considered for the decay of the Higgs boson. 
The $A$ boson samples are generated for both production mechanisms 
and are also employed for modelling $H$ and $h$ production.
The differences between CP-even and CP-odd eigenstates are negligible for this analysis.
The signal modelling for a given combination of $\MA$ and $\tan\beta$
takes into account all three Higgs bosons $h$, $H$ and $A$,
by adding the $A$ boson mass samples corresponding to $m_h$, $m_H$ and $\MA$ according to their
production cross-sections and masses
in the $\mhmax,\, \mu > 0$ MSSM benchmark scenario.

For the \tautau{} decay mode, 15 Monte Carlo samples are generated with Higgs boson masses 
in the range of \unit[90]{\GEVcc} to \unit[500]{\GEVcc} and $\tan\beta = 20$. These
are scaled to the appropriate cross-sections for other $\tan\beta$ values. 
The simulated signal samples with $\MA$ closest to the computed mass of $H$ and $h$ are used for the $H$ and $h$ bosons.
The increase in the Higgs boson natural width with $\tan\beta$, of the order of \unit[1]{\GEVcc} in the range considered, 
is negligible compared to the experimental mass resolution in this channel, which is always above \unit[10]{\GEVcc}.

For the \mumu{} decay mode, seven samples are generated with Higgs boson masses in
the range of \unit[110]{\GEVcc} to \unit[300]{\GEVcc} and $\tan\beta = 40$.  Additionally, to study the $\tan\beta$ 
dependence of the width of the resonance, signal samples are generated for both
production modes for $\MA=\unit[150]{\GEVcc}$ and \unit[250]{\GEVcc}, each at $\tan\beta = 20$ and $\tan\beta = 60$.  
Since the mass resolution is better in this channel, signal distributions are obtained using an interpolation procedure
 for different intermediate $\MA$--$\tan\beta$ values, as  described in \secref{sec:mumu_channel}.

The generated Monte Carlo samples for the \Htotautau{} decay modes are passed through the full GEANT4~\cite{ATLASSIM, Geant4} detector simulation, while the samples for the \Htomumu{} decay mode are passed through the full GEANT4 detector simulation or the ``fast'' simulation, ATLFAST-II~\cite{ATLASSIM}, of the ATLAS detector.

\paragraph{Background processes:}

The production of $W$ and $Z/\gamma^*$ bosons in association with jets is simulated 
using the ALPGEN~\cite{Alpgen} and PYTHIA~\cite{Pythia} generators. 
PYTHIA is also used for the simulation of $\bbbar$ production, but through an interface, which ensures that the simulation 
is in agreement with $b$-quark production data \cite{Baranov,pythiaB2}.
The $\ttbar$ production process is generated with MC@NLO~\cite{MCatNLO} and POWHEG~\cite{Powheg1,Powheg2}.  MC@NLO is used for the generation of electroweak diboson~($WW$, $WZ$, $ZZ$) samples.  Single-top production through the $s$- and $t$-channels, and in association with $W$ bosons, is generated using AcerMC~\cite{AcerMC}.
For all event samples described above, parton showers and hadronisation are simulated with HERWIG~\cite{Herwig} 
and the activity of the underlying event with JIMMY~\cite{JIMMY}.
The loop-induced $gg \to WW$ processes are generated using gg2WW~\cite{GG2WW}.
The following parton distribution function sets are used: CT10~\cite{CT10} for MC@NLO,
CTEQ6L1~\cite{CTEQ6L1} for ALPGEN and modified leading-order MRST2007~\cite{MRST2007lomod} for PYTHIA samples.

Decays of \atau{} leptons are simulated
using either SHERPA or TAUOLA~\cite{TAUOLA}.  
Initial-state and final-state radiation of photons is simulated using
either PHOTOS~\cite{PHOTOS} or, for the samples generated with SHERPA, PHOTONS++, which is a part of SHERPA. 
The \Ztotautau{} background processes are modelled with a $\atau$-embedded $\Ztomumu$ data sample 
described in \secref{sec:tautau_channel}. All generated Monte Carlo background samples are passed
through the full GEANT4 simulation of the ATLAS detector.

The signal and background samples are reconstructed with the same software as used for data. To take into account the presence of multiple interactions occurring in the same and neighbouring bunch crossings (referred to as pile-up), simulated minimum bias events are added to the
hard process in each generated event. Prior to the analysis, simulated events are re-weighted in order to match the distribution of the number of pile-up interactions per bunch crossing in the data.

%% file: objects.tex
\section{Physics object reconstruction}
\label{sec:objects}

An electron candidate is formed from energy deposits in the electromagnetic calorimeter associated with a 
charged particle track measured in the inner detector. 
Electrons are selected if they have a transverse energy $\ET > \unit[15]{\GeV}$, lie within
$|\eta|<2.47$, but outside  the transition region between the barrel and end-cap 
calorimeters ($1.37<|\eta|<1.52$), and 
meet quality requirements based on the expected shower shape~\cite{egammapaper}. 

A muon candidate is formed from a high-quality track measured in the inner detector matched to hits
in the muon spectrometer~\cite{MuonEfficiencyConfNote}. 
Muons are required to have a transverse momentum $\pt > \unit[10]{\GEVc}$ and to lie within $|\eta| < 2.5$. 
In addition, the point of closest approach of the inner detector track must be no further than
\unit[1]{cm} from the primary vertex\footnote{The primary vertex is defined as the vertex with the largest
$\Sigma \pt^{2}$ of the associated tracks.}, 
as measured along the $z$-axis.
This requirement reduces the contamination due to cosmic-ray muons and beam-induced 
backgrounds. 

Identified electrons and muons are isolated if there is little additional activity 
in the inner detector and the calorimeter around the electron or muon.
The scalar sum of the transverse momenta of all tracks from the same vertex as the lepton,
with \pt\ above \unit[1]{\GEVc} and located within a cone with radius parameter\footnote{$\Delta R = \sqrt{\left(\Delta\eta\right)^2 + \left(\Delta\phi\right)^2}$, where $\Delta\eta$ is the difference in pseudorapidity of the two objects in question and $\Delta\phi$ is the difference between their azimuthal angles.} 
$\Delta R=0.4$
 around the lepton direction,
must be less than \percent{6} of the lepton momentum for the $\tautau$ channels, or less than \percent{10} for the $\mumu$ channels. 
The sum excludes the track associated to the lepton itself. 
In addition, for the $\tautau$ channel, the remaining calorimetric energy within a cone of radius parameter
$\Delta R =0.2$ around the lepton direction must be less than \percent{8} (\percent{4}) of the electron 
transverse energy (muon transverse momentum). The remaining energy excludes that associated to the lepton itself 
and an average correction due to pile-up is made.

Reconstructed electrons and muons are used to identify the leptonic decays of $\tau$ leptons ($\tlep$). The decay of a $\tau$ lepton into an electron (muon) and neutrinos is denoted by $\telec$ ($\tmuon$).

Jets are reconstructed using the anti-$k_{t}$ algorithm~\cite{AntiKT} with a 
radius parameter $R=0.4$, taking three-dimensional 
noise-suppressed clusters of calorimeter-cell energy deposits~\cite{TopoClusterAlgo}
as input.
The jet energy is calibrated using a combination of test-beam results, simulation 
and in situ measurements~\cite{ATLASJETEnergyScale}.
Jets must satisfy $E_T > \unit[20]{\GeV}$ and $|\eta| < 4.5$. 
Rare events containing a jet with associated out-of-time activity or calorimeter noise are discarded.
Tracks are classified as associated tracks if they lie within $\Delta R < 0.4$ of the jet axis.   
To reduce the pile-up effect, the scalar sum of the transverse momenta of the associated tracks 
matched to the primary vertex must be at least \percent{75} of the jet transverse momentum measured in the calorimeter.
A multivariate algorithm based on a neural network is used in this analysis
to tag jets, reconstructed within $\left|\eta\right|<2.5$, that are associated with the hadronisation of $b$-quarks.
The neural network makes use of the impact parameter of associated tracks and  
the reconstruction of $b$- and $c$-hadron decay vertices inside the jet~\cite{BtaggingEfficiency}. 
The \bjet{} identification has an efficiency of about \percent{70} in $\ttbar$ events, unless
otherwise stated.
The corresponding rejection factors are about 5 for jets containing charm hadrons and about 130 for light-quark or gluon
jets.

Hadronic decays of $\tau$ leptons (\thad)  are characterised by the presence of one, three, or in rare cases, five  
charged hadrons accompanied by a neutrino and possibly neutral hadrons, resulting in 
a collimated shower profile in the calorimeters with only a few nearby tracks. The visible $\tau$ 
decay products are  reconstructed in the same way as jets,
but are calibrated separately to account for the different calorimeter response  compared to jets.
Information on the collimation, isolation, and shower profile
is combined into a boosted-decision-tree discriminant to reject backgrounds from jets~\cite{ATLASTauIDNew}. 
In this analysis, three selections are used ---``loose'', ``medium'' and ``tight''--- with identification efficiency 
of about \percent{60}, \percent{45} and \percent{35}, respectively.
The rejection factor against jets varies from about 20 for the loose selection 
to about 300 for the tight selection.
A \thad{} candidate must lie within $|\eta| < 2.5$, have a transverse momentum greater than \unit[20]{\GEVc}, 
one or three associated tracks (with $\pt>\unit[1]{\GEVc}$), and a total charge of $\pm 1$.  
Dedicated electron and muon veto algorithms are applied to each $\thad$ candidate.

When different objects selected according to 
the above criteria overlap with each other geometrically (within 
$\Delta R = 0.2$), only one of them is 
considered for further analysis. The overlap is resolved by selecting
muon candidates, 
electron candidates, $\thad$ candidates and jet candidates in this order of priority.

The magnitude and direction of the missing transverse momentum, $\met$, is reconstructed including 
contributions from muons and energy deposits in the calorimeters~\cite{ATLASmetNEW}. 
Clusters of calorimeter-cell energy deposits belonging to jets, $\thad$ candidates, electrons, and photons, as well as cells that are not associated to any object, are treated separately in the $\met$ calculation.
The contributions of muons to $\met$ are calculated differently for isolated and non-isolated muons, to account for the energy deposited by muons in the calorimeters.

%% file: mumu_channel.tex
\section{The \mumu{} decay channel} 
\label{sec:mumu_channel}

\paragraph{Signal topology and event selection:}
The signature of the \Amumu{} decay is  a pair of isolated muons with high transverse momenta and opposite charge.
In the $b$-associated production mode, the final state can be further characterised by the presence of one or two low-$E_T$ \bjets . 
The missing transverse momentum is expected to be small and on the order of the resolution of the \met{} measurement.
The \mumu{} decay channel search is complicated by a small branching ratio and considerable background rates.

Events considered in the $\mumu$ analysis must pass a single-muon trigger with a transverse momentum threshold of \unit[18]{\GEVc}.
At least one reconstructed muon is required to be matched to the $\eta$--$\phi$ region of the trigger object and to have $\pT > \unit[20]{\GEVc}$.
At least one additional muon of opposite charge and with $\pT > \unit[15]{\GEVc}$ is required. A muon pair is formed using the two highest-$\pt$ muons of opposite charge. 
This muon pair is required to have an invariant mass greater than \unit[70]{\GEVcc}.
In addition, events are required to have $\met < \unit[40]{\GEVc}$.
All muons considered here must be isolated, as defined in section~\ref{sec:objects}.

\begin{figure}[t]
\centering
\subfigure{\includegraphics[width=0.49\textwidth]{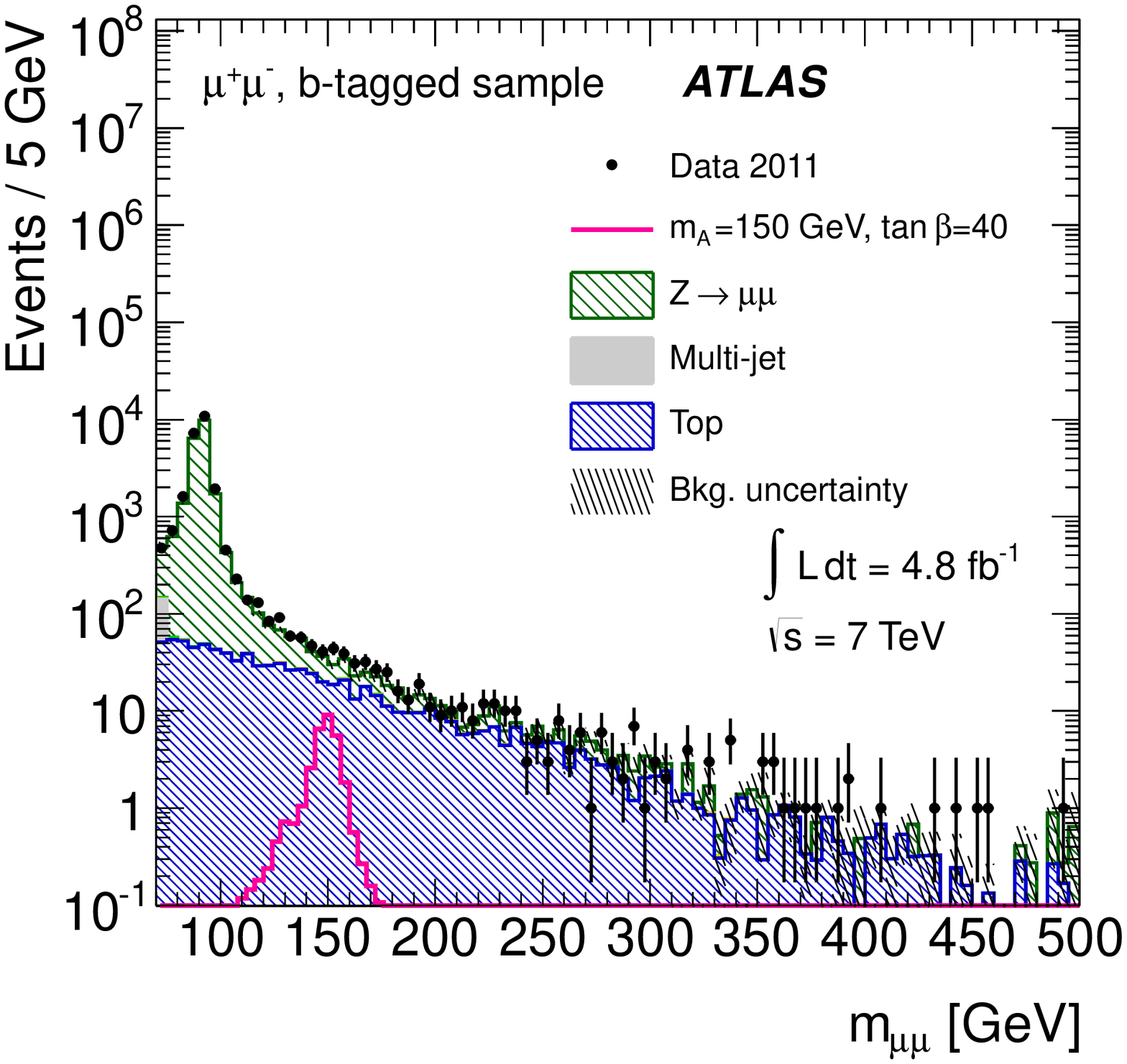}}
\subfigure{\includegraphics[width=0.49\textwidth]{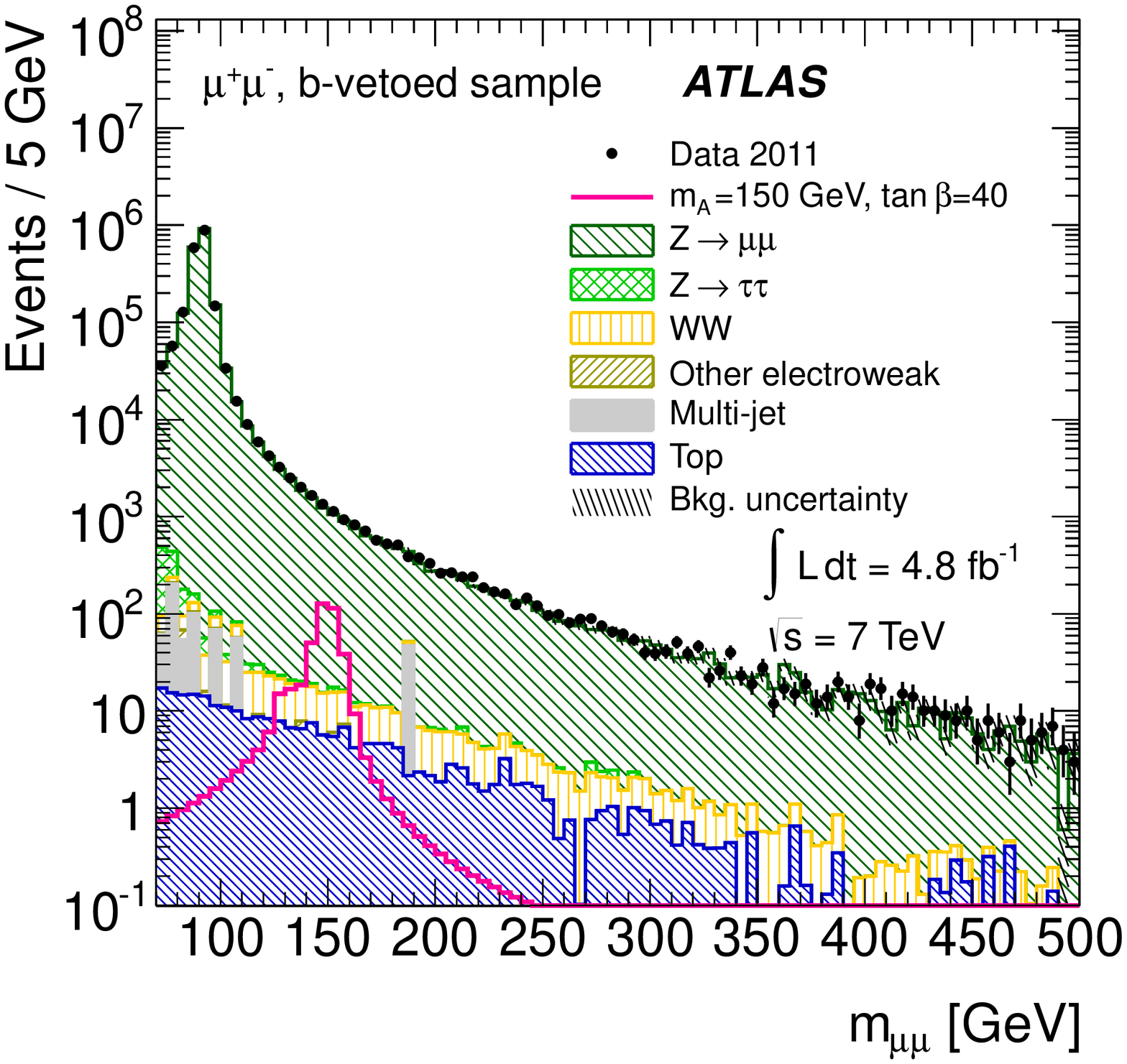}}
\caption{
The invariant mass distribution of the two muons of the \Htomumu{} search for the $b$-tagged (left-hand side) and the $b$-vetoed samples (right-hand
side). The data are compared to the background expectation and a hypothetical MSSM signal ($\MA=\unit[150]{\GEVcc}$, $\tan\beta=40$). Simulated backgrounds are shown for illustration purposes. The background uncertainties shown here are statistical in nature due to the finite number of simulated events.
The contributions of the backgrounds \Ztoee, \Wjets{} and those of all diboson production processes but $WW$ production are combined and labelled ``Other electroweak''.
} \label{fig:mumu_inv_mass}
\end{figure}

The large background due to  $Z/\gamma^{*}$ production can be reduced by requiring
that the event contains at least one identified \bjet.
Events satisfying this requirement are included 
in the $b$-tagged sample, whereas events failing it are included in the $b$-vetoed sample.
The $\mumu$ invariant mass distribution, $m_{\mu\mu}$, is shown separately for the $b$-tagged
and the  $b$-vetoed samples in \figref{fig:mumu_inv_mass}. For illustration purposes only, the
distributions of simulated backgrounds and an assumed MSSM neutral Higgs boson signal with
$\MA=\unit[150]{\GEVcc}$ and $\tan\beta = 40$ are shown in the same \figureword.
A hypothetical signal would be present as narrow peaks on top of the high-mass tail of the $Z$ boson superimposed on a continuous 
contribution from non-resonant backgrounds such as \ttbar{}. The \Zgstar{} process contributes to the total background 
with a relative fraction of about \percent{99} (\percent{51}) for the $b$-vetoed ($b$-tagged) sample for events in the $m_{\mu\mu}$ range of \unit[$110$]{\GEVcc} to \unit[$300$]{\GEVcc}, 
which is most relevant to the Higgs boson searches in this channel.
In the $b$-vetoed sample the remaining non-resonant background is composed of \ttbar{}, \WW{} and \bbbar{} events
while $\ttbar{}$ events dominate the non-resonant background in the $b$-tagged sample.

\paragraph{Background modelling:}
The background in the $\mumu$ channel is estimated from data.
By scanning over the $\mumu$ invariant mass distribution, local sideband fits provide the expected background estimate in the mass region of interest.
To this end, a parameterisation of the background shape is fitted to the $\mumu$ invariant mass distribution. 
Search windows are defined around each of the neutral Higgs bosons and are excluded from the fit.
This results in one or two windows due to the mass degeneracy among the Higgs bosons.
The widths of the search windows are motivated by the expected signal width for each point in the scanned $\MA$--$\tb$ grid and account for asymmetries in the signal invariant mass distribution.
The upper and lower boundaries of the search windows are defined by the $m_{\mu\mu}$ values where the 
cross-section predictions of the signal model are \percent{10} of their maximum.
The lower and upper outer boundaries of the sidebands vary between 98--$\unit[118]{\GEVcc}$ and 160--$\unit[400]{\GEVcc}$, 
respectively, depending on  $m_A$ and $m_h$.

The parameterisation of the $\mumu$ invariant mass distribution, $f_B(x)$, is given by 
\begin{equation} \label{eq:bkg_fit}
f_{B}\left(x\right.\left|N_{B},A,B,m_{Z},\Gamma_{Z},\sigma\right) = N_{B} \cdot \left[ f_{Z}\left(x\right.\left|A,B,m_{Z},\Gamma_{Z}\right) \otimes \mathcal{F}_{\text{G}}\left(x\right.\left|0,\sigma\right)  \right],
\end{equation}
where $x$ is the  invariant mass, $\otimes$ the convolution operator and $\mathcal{F}_{\text{G}}\left(x\right.\left|0,\sigma\right)$ the Gaussian distribution with variable $x$, mean 0 and variance $\sigma^2$.
The function $f_Z$ describing the $Z/\gamma^{*}$ production is
\begin{equation} \label{eq:bw}
f_{Z}\left(x\right.\left|A,B,m_{Z},\Gamma_{Z}\right) = A\frac{1}{x^{2}} + B\frac{x^{2}-m_{Z}^{2}}{\left( x^{2}- m_{Z}^{2} \right)^{2} + m_{Z}^{2}\Gamma_{Z}^{2}} + 
\frac{x^{2}}{\left( x^{2}- m_{Z}^{2} \right)^{2} + m_{Z}^{2}\Gamma_{Z}^{2}}.
\end{equation}
This is convolved with the Gaussian distribution accounting for the finite mass resolution.
The function $f_{Z}$ is a simplification of the pure $\gamma^{*}$ and $Z$ propagators, including $Z$--$\gamma^{*}$ interference contributing to the process 
$\qqbar \ra Z/\gamma^{*} \ra \mumu$, and hence in principle only describes the background from $Z/\gamma^{*}$ production. 
The parameterisation $f_B$ is found to be a good approximation of the shape of the total $\mumu$ background 
even in the $b$-tagged sample, which has non-negligible contributions from physics
processes other than $Z/\gamma^{*}$ production.

In total, the fit function, $f_{B}$, has six parameters.
The natural width of the $Z$ boson, $\Gamma_{Z}$, is fixed to $\Gamma_Z = \unit[2.50]{\GEVcc}$, whereas the remaining parameters are unconstrained. 
Parameter $N_{B}$ describes the total normalisation of the curve and parameters $A$ and $B$ represent the relative normalisations of the $\gamma^{*}$ and
$Z$--$\gamma^{*}$ contributions with respect to the $Z$ term. Finally,
$m_Z$ represents the mass of the $Z$ boson and the parameter $\sigma$ specifies the mean $\mumu$ pair mass resolution.

For every point on the $\MA$--$\tb$ grid, a binned likelihood fit of $f_B$ to the data is performed to estimate the  five unconstrained 
parameters and consequently the total background estimate.

The background model is validated with $\chi^{2}$-based goodness-of-fit studies. 
In addition, the background model is extended by  polynomials of different orders to test if additional degrees of 
freedom change the goodness of the fit, 
which would hint at problems in the shape modelling.
Further validation of the capability of the model to describe the shape of the data is performed by varying the fit ranges for certain 
mass points and accounting for the fit residuals.
The goodness-of-fit studies confirm a good background modelling for both 
the $b$-vetoed and the $b$-tagged sample.
The uncertainty on the background estimate is obtained from a variation of the fitted background function within its
\percent{68} confidence level (CL) uncertainty band.
This results in an uncertainty of \percent{5} (\percent{2}) on the expected background yield for the $b$-tagged ($b$-vetoed) sample.

\paragraph{Signal modelling:}
The $\Amumu$ signal is expected to appear as narrow peaks in the $\mumu$ invariant mass distribution, as depicted in \figref{fig:mumu_inv_mass}.
The resolution in the relevant mass range is typically \percent{$2.5$} to \percent{$3$}, and numerous mass points are needed for a complete mass scan. 
In addition, the influence of $\tb$ on the reconstructed width of the signal invariant mass distribution needs to be taken into account.
The natural widths of the MSSM neutral Higgs bosons increase with $\tb$. 
The reconstructed width can be sensitive to this variation because of the good experimental mass resolution.

To interpolate between the different signal samples obtained from a 
limited number of simulated signal masses, the signal $\mumu$ invariant 
mass distribution is parameterised with
\begin{align} \label{eq:signal_fit}
	f_{S}\left(x\right.\left|N_{S},m,\Gamma,\sigma,c,\varsigma\right) = N_{S} &\left[ \frac{1}{\left[ x^{2}-m^{2} \right]^{2} + m^{2}\Gamma^{2}} \otimes \mathcal{F}_{\text{G}}\left(x\right.\left|0,\sigma\right)\right.\nonumber\\ &\left. \hphantom{\left[\right.}\vphantom{\frac{1}{\left[ x^{2}-m^{2} \right]^{2} + m^{2}\Gamma^{2}}} + c\cdot \mathcal{F}_{\text{L}}\left(-x\right.\left|m,\varsigma\right) \right],
\end{align}
where $x$ represents the $\mumu$ invariant mass.
The parameterisation consists of a Breit--Wigner function describing the signal peak convolved 
with a Gaussian distribution, $\mathcal{F}_{\text{G}}$, accounting for the finite mass resolution
and a Landau function, $\mathcal{F}_{\text{L}}$, with low-mass tail which
models the asymmetric part of the signal invariant mass distribution.

The function $f_S$ is characterised by six parameters. The width of the Breit--Wigner function, $\Gamma$, is fixed to the theoretical 
predictions calculated with FeynHiggs~\cite{feynhiggs}. The remaining five parameters are  unconstrained.
The overall normalisation parameter is $N_{S}$ and $c$ specifies the relative normalisation of the Landau function with respect to the Breit--Wigner function.
Parameter $m$ specifies the mean of the Breit--Wigner and the Landau distributions, $\sigma$ determines the width of the Gaussian distribution and $\varsigma$ 
represents the scale parameter of the Landau function.

The function $f_{S}$ is fitted to each signal sample available from simulation. The  signal model is validated with Kolmogorov-Smirnov- and $\chi^2$-based  goodness-of-fit tests proving a good description of the simulated signal  $\mu^{+}\mu^{-}$ invariant mass distributions.
Each fit results in a set of fitted parameters, $(N_{S},m,\sigma,c,\varsigma)$, depending on the point in the $\MA$--$\tan\beta$ plane.
The dependence of this set on $\MA$ and $\tb$ is parameterised with polynomials of different orders.
The resulting polynomials 
provide a set of parameters which in addition to the predicted natural width, $\Gamma$, fully define the normalised 
probability density function for an arbitrary point in the $\MA$--$\tb$ plane.

This procedure is used to generate invariant mass distributions for signal masses from \unit[$120$]{\GEVcc} to \unit[$150$]{\GEVcc} in \unit[5]{\GEVcc} steps and from \unit[$150$]{\GEVcc} to \unit[$300$]{\GEVcc} 
in \unit[10]{\GEVcc} steps, as well as for $\tan\beta$ values from $5$ to $70$ in steps of $3$ or $5$.
Higgs boson masses below \unit[120]{\GEVcc} were not considered because the background model does not provide precise estimates in the mass region close to the $Z$ boson peak.
For both the $b$-tagged and the $b$-vetoed samples the interpolated and normalised probability
density functions are obtained separately for the Higgs boson production from
gluon-fusion and in association with \bquarks .
As for the background, the uncertainty on the signal prediction from the fit is obtained from 
its \percent{68} CL uncertainty band.
The resulting uncertainty is estimated to be \percent{10} to \percent{20} of the signal event yield.

\paragraph{Results:}

\begin{figure}[t]
\centering
\subfigure{\includegraphics[width=0.49\textwidth]{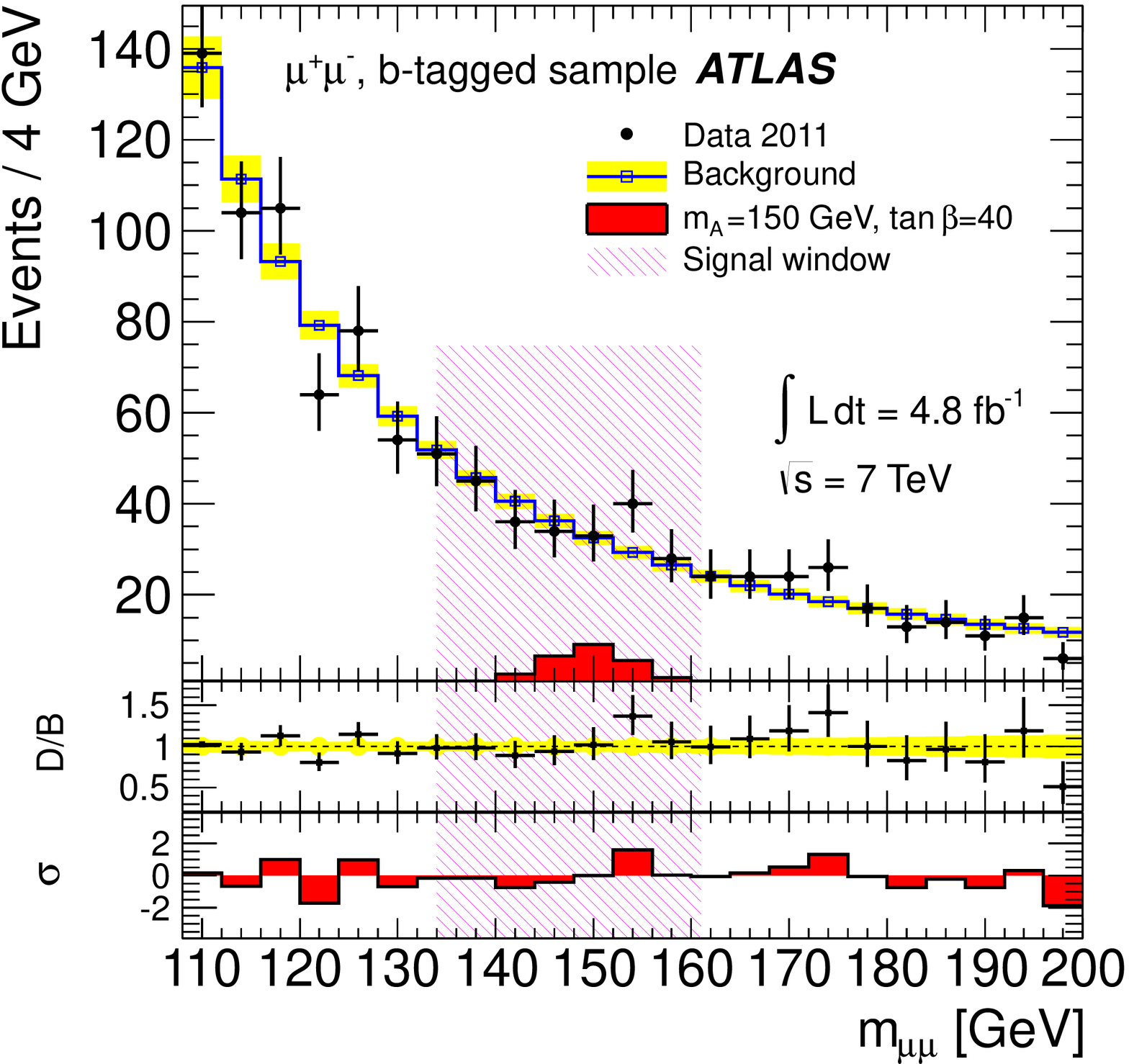}}
\subfigure{\includegraphics[width=0.49\textwidth]{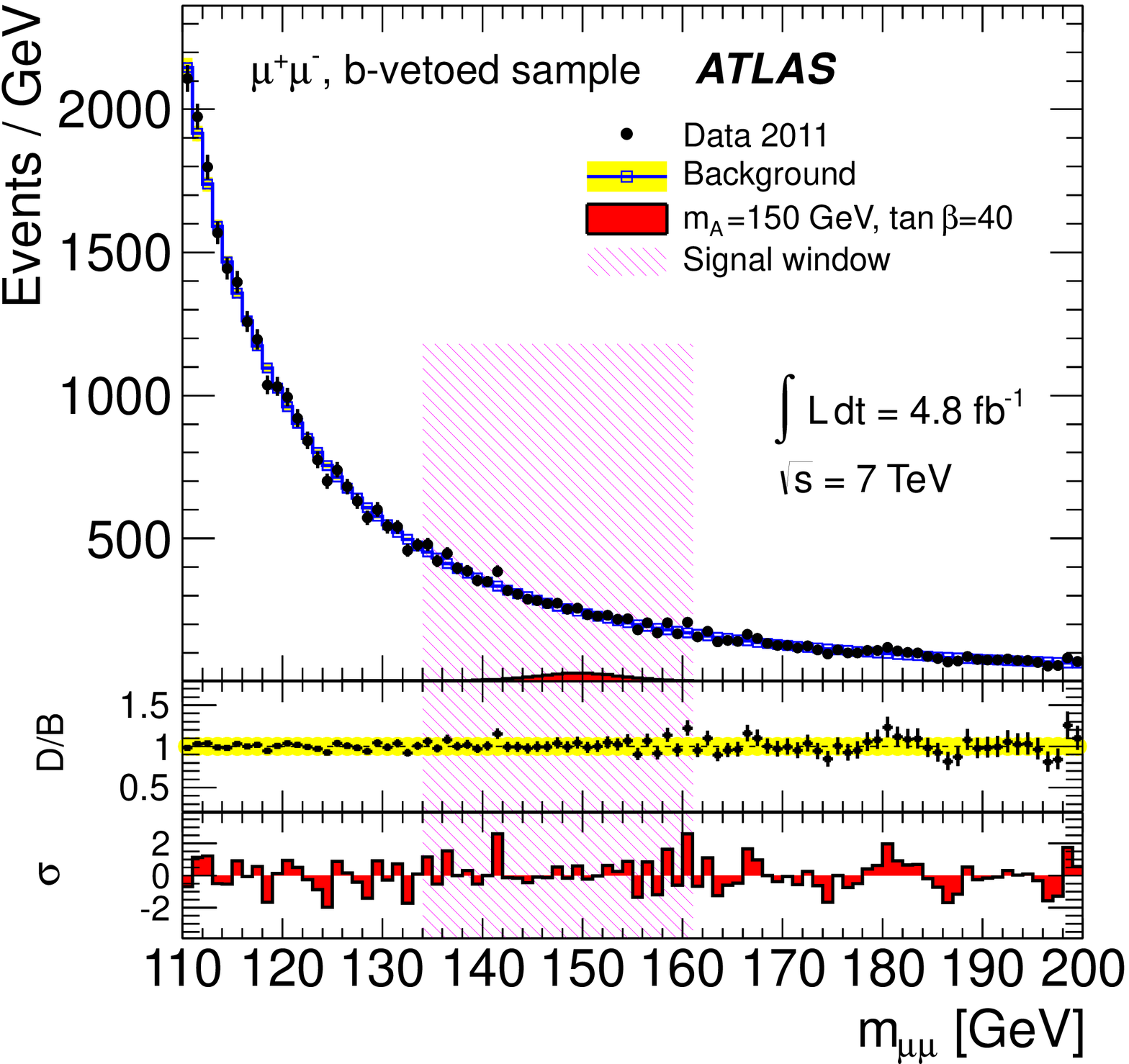}}
\caption{Invariant mass distribution of data and predicted background from sideband fits to the data
shown for the signal mass point at $\MA=\unit[150]{\GEVcc}$ and $\tb=40$ for the $b$-tagged (left-hand side) and the $b$-vetoed samples (right-hand side) of the
\Htomumu{} final state. The ratio of the data to the predicted background, labelled $D/B$, and the
bin-by-bin significances of the deviations of the data from the background prediction,
labelled $\sigma$, are shown beneath.}
\label{fig:mumu_background}
\end{figure}

\Figref{fig:mumu_background} compares the data with the background estimate predicted from sideband fits in both the $b$-tagged 
and $b$-vetoed samples for the signal mass point $\MA=\unit[150]{\GEVcc}$ and $\tb=40$.
The data fluctuate around the background prediction leading to local bin-by-bin significances that are  typically less than $2\,\sigma$.
\Tabref{tab:mumu_fit} shows the number of observed events in the fit range around the mass point $\MA=\unit[150]{\GEVcc}$ compared to 
the number of background events predicted by the sideband fits. 
The observed numbers of events are compatible with the expected yield from Standard Model processes within the uncertainties.

\begin{table}
\centering
\input{./mumu_fit.tex}
\caption{
Observed number of data events and the expected number of signal and background events in the \Htomumu~channel for one of the considered signal mass points. The number of background events is
predicted from sideband fits to the $\mumu$ invariant mass distribution in the fit range around the signal mass point $\MA=\unit[150]{\GEVcc}$ for both the \Amumu{} $b$-vetoed and $b$-tagged samples. The number of expected signal events produced in gluon-fusion or in association with \bquarks\ is shown for $\tb=40$.
The quoted uncertainty for the background estimate is the statistical uncertainty obtained from the fit.
For the signal estimate, the uncertainty from the fit is quoted first and 
then separately the uncertainty from other sources discussed in \secref{sec:systematics}.
}
\label{tab:mumu_fit}
\end{table}

%% file: mumu_fit.tex
\renewcommand\arraystretch{1.2}
\begin{tabular}{lc@{\extracolsep{3pt}}lcc@{\extracolsep{3pt}}l}
\hline\hline
					& \multicolumn{2}{c}{$b$-tagged sample} & & \multicolumn{2}{c}{$b$-vetoed sample}	\\
\hline
Mass Point				& \multicolumn{5}{c}{$m_A=\unit[150]{\GeV}$}	\\
Fit Range				& \multicolumn{5}{c}{$\unit[ 110\textrm{--}200 ]{\GeV}$}	\\
\hline
Background                                    & \multicolumn{2}{c}{$980  \pm 50$}    &  & \multicolumn{2}{l}{$35900 \pm 600$} \\
\hline
%Signal & & && &\\ $m_A=\unit[150]{GeV}, \tan\beta=40$ & & && &\\
%\multicolumn{6}{l}{Signal $m_A=\unit[150]{GeV}, \tan\beta=40$}\\
Signal $m_A=\unit[150]{GeV}, \tan\beta=40$ & & && & \\
%\multicolumn{6}{c}{Signal expectation for $m_A = \unit[150]{\GeV}, \tan\beta = 40$}\\
%\multicolumn{6}{c}{Signal $m_A = \unit[150]{\GeV}, \tan\beta = 40$}\\
$b\bar{b}$(\Higgses$\rightarrow\mu\mu$)       & $28  \pm 2\hphantom{.0}$      & \err{3}{4} &  & $271 \pm 22$ & \err{31}{40} \\
$gg\rightarrow$\Higgses$\rightarrow\mu\mu$    & $ 2.3 \pm 0.3$  & $\pm 0.4$     &  & $141 \pm 10$ & \err{22}{20} \\
\hline
Data                                          & \multicolumn{2}{c}{$985$} & & \multicolumn{2}{c}{$36044$} \\
\hline\hline
\end{tabular}
\renewcommand\arraystretch{1.}

%% file: tautau_channel.tex
\section{The $\tautau$ decay channel} \label{sec:tautau_channel}

The \Htotautau{} decay mode is analysed in several categories according to the $\tau$ lepton decay
final-state combinations.
The four decay modes considered here are: $\telec\tmuon$~(6\%), $\telec\thad$~(23\%), $\tmuon\thad$~(23\%) and $\thad\thad$~(42\%),
where  percentages in the parentheses denote the corresponding branching ratios.
The combination of $\telec\thad$ and $\tmuon\thad$ is referred to as $\tlep\thad$.

\subsection{Common background estimation and mass reconstruction techniques}

\input{embedding}

\input{jettotaufakes}

\input{ABCDmethod}

\input{MMC}

\input{leplep}

\input{lephad}

\input{hadhad}

%% file: embedding.tex
\paragraph{$\boldsymbol{\tau}$-embedded $\boldsymbol{\Ztomumu{}}$ data:} \label{sec:embedding}

\Ztotautau{} events form a largely irreducible background to the Higgs boson signal in all  final states. 
It is not possible to select a \Ztotautau{} control sample which is Higgs boson signal-free. However, \Ztomumu{} events can be selected in data with high purity and without significant signal contamination. 
Furthermore, the event topology and kinematics are, apart from the $\tau$ lepton decays and the  different masses
of $\tau$ leptons and muons, identical to those of \Ztotautau{} events. Therefore \Ztomumu{}
events are selected in data and modified using a $\tau$-embedding
technique, in which muons are replaced by simulated \tauleptons. The hits of the muon tracks and the
associated calorimeter cells in a cone with radius parameter $\Delta R=0.1$ around the muon direction are removed
from the data event and replaced by the detector response from a simulated \Ztotautau{} event with the same kinematics. The
event reconstruction is performed on the resulting hybrid event.  Only the $\tau$ decays and
their detector response are taken from the simulation, whereas the underlying event kinematics
and the associated jets are taken from the data event. The procedure treats consistently the effect
of $\tau$ polarisation and spin correlations. 
The event yield of the embedded sample after the selection of the $\tau$ decay products is normalised
to the corresponding event yield obtained in a simulated \Ztotautau{} sample.
This procedure has been validated as described in \refscite{ATLASLimit,SMHtautau2011}.

Systematic uncertainties on the normalisation and shape of the embedded sample
are derived by propagating variations of the \Ztomumu{} event selection and the muon energy
subtraction procedure through the $\tau$-embedding process. 
Additional uncertainties are assigned 
due to the use of the \Ztotautau{} cross-section and Monte Carlo acceptance prediction
in determining the  $\tau$-embedded \Ztomumu{} sample normalisation.
These theoretical uncertainties are described in \secref{sec:systematics}.

%% file: jettotaufakes.tex
\paragraph{Jets misidentified as hadronic $\boldsymbol{\tau}$ decays:} \label{sec:jettotaufakes}

A fraction of jets originating from quarks or gluons
are misidentified as \thad{} candidates.
It has been shown in \refcite{AtlasZtautau} %\refcite{TauFR}
that this misidentification fraction is higher in
simulated samples than in data. To account for this difference, 
the Monte Carlo background estimate is
corrected based on control samples.
Details are presented for each decay channel separately.

%% file: ABCDmethod.tex
\paragraph{The $\boldsymbol{ABCD}$ background estimation method:} \label{sec:ABCD_description}

The estimation of the background from  multi-jet processes is done from data
using the $ABCD$ method for all \tautau{} channels. Two uncorrelated variables are chosen to define four
data regions, named $A$, $B$, $C$ and $D$, such that one variable separates $A$ and $B$ from $C$ and $D$, while the other separates $A$ and $C$ from $B$ and $D$.
The signal region is labelled $A$,
and the other regions are dominated by background from  multi-jet processes.
An estimate of the background from these processes in the signal region, $n_A$, is:
\begin{equation}
	\label{eq:ABCD_formula}
	n_A = n_B \times \frac{n_C}{n_D} \equiv n_{B} \times r_{C/D},
\end{equation}
where $n_B$, $n_C$ and $n_D$ denote the populations of regions $B$, $C$ and $D$, respectively.
The populations of the $B$, $C$ and $D$ regions may need to be corrected by subtracting
the estimated number of events that come from processes other than  multi-jet production.
This estimate is generally obtained from  simulation.

%% file: MMC.tex
\paragraph{$\boldsymbol{\tautau{}}$ mass reconstruction:} \label{MMC}

The invariant mass of the $\tautau$ pair cannot be reconstructed
directly due to the presence of neutrinos from the $\tau$ lepton decays.
Therefore, a technique known as the Missing Mass Calculator (MMC)
is used to reconstruct the Higgs boson candidate  mass~\cite{MMCpaper}.
This algorithm assumes that the missing transverse momentum is due
entirely to the neutrinos, and performs a scan over the angles between 
the neutrinos and the visible $\tau$ lepton decay products. For leptonic $\tau$ 
decays, the scan also includes the invariant mass of the two neutrinos.
At each point, the \tautau{} invariant mass is calculated, and the most
likely value is chosen by weighting each solution according to
probability density functions that are derived from simulated $\tau$ lepton decays.
This method provides a \percent{$13$} to \percent{$20$} resolution in the invariant mass, with
an efficiency of \percent{99} for the scan to find a solution.

%% file: leplep.tex
\subsection{The \Htotautauemu\ decay channel} \label{sec:leplep}

\paragraph{Signal topology and event selection:}

Events in this channel must satisfy either a single-electron, single-muon or combined electron--muon trigger.
The single-lepton triggers have \pt{} thresholds of \unit[20]{\GEVc} or \unit[22]{\GEVc} for electrons, 
depending on the run period,  and \unit[18]{\GEVc} for muons, while
the combined trigger has a threshold of \unit[10]{\GEVc} for the electron and \unit[6]{\GEVc} for the muon.
Exactly one isolated electron and one isolated muon of opposite electric charge are required, and
the lepton pair must have an invariant mass exceeding \unit[30]{\GEVcc}. 
The $\pt$ thresholds are  \unit[15]{\GEVc} for electrons and \unit[10]{\GEVc} for muons in cases where the
event is selected by the combined electron--muon trigger.
These thresholds are  raised to \unit[24]{\GEVc} for electrons and \unit[20]{\GEVc} for muons 
in cases where the event is selected by a single-lepton trigger only.

The event sample is then split according to its jet flavour content.
Events containing exactly one identified \bjet\ are included in the $b$-tagged sample.
This is based on a \bjet{} identification criterion with \percent{75} efficiency in \ttbar{} events.
Events without an identified \bjet{} are included in the $b$-vetoed sample.
The scalar sum of the lepton transverse momenta and missing transverse momentum is required to fulfil 
the following condition to reduce top quark and diboson backgrounds:
 $\HTlep < \unit[125]{\GEVc}$ ($<\unit[150]{\GEVc}$) for the $b$-tagged ($b$-vetoed) sample.
In order to further suppress these backgrounds and
\Wtolnu{} events, the opening angle between the
two lepton candidates in the transverse plane must satisfy
 the condition $\deltaPhi>2.0$ ($>1.6$) for the $b$-tagged ($b$-vetoed) sample.
In addition, the combination of the transverse opening angles between the lepton directions
and the direction of \MET{} is required to satisfy the condition $\sumcos > -0.2$ ($>-0.4$) for the $b$-tagged ($b$-vetoed) sample. 
Finally, the scalar sum of the transverse energies of all jets, $\HT$,
is restricted to be below \unit[100]{\GeV} in the $b$-tagged sample to further 
suppress backgrounds containing a higher multiplicity of jets, or jets with higher transverse momenta, than expected from the signal processes. 
Jets with $|\eta| < 4.5$ and $E_T > \unit[20]{\GeV}$ are used to calculate the value of $\HT$.

\paragraph{Estimation of the $\boldsymbol{\Ztotautau{}}$ background:}

The \Ztotautau{} background is estimated by using the $\tau$-embedded \Ztomumu{} event
sample outlined in \secref{sec:embedding}.
The use of multiple triggers with different \pt{} thresholds
has an effect on the lepton transverse momentum spectra, which is accurately reproduced by the
trigger simulation. However, in the $\tau$-embedded \Ztomumu{} data there is no simulation of the
trigger response for the decay products of the $\tau$ leptons. This has an impact on the MMC mass
distribution in the $\tau$-embedded \Ztomumu{} data, which is comparable to the statistical
uncertainty in the $b$-vetoed sample, and negligible in the $b$-tagged sample.
For this reason the trigger selection is emulated for the $b$-vetoed sample 
such that the trigger effect is adequately described. This emulation is based on the \pt-dependent trigger
efficiencies obtained from data.

\paragraph{Estimation of the $\boldsymbol{\ttbar{}}$ background:} \label{sec:leplep_btag_ttbarestimation}
The contribution of \ttbar{} production 
is extrapolated from control regions which have  purities of \percent{90} ($b$-tagged sample)
and \percent{96} ($b$-vetoed sample). 
The selection criteria for these control regions are
identical to the respective signal regions with two exceptions:
at least two identified \bjets\ are required, 
and the selection $\HT < \unit[100]{\GeV}$ is not applied. 
The multi-jet contributions to these control regions are estimated from data with the $ABCD$ method; 
the other non-\ttbar{} contributions are taken from simulation. 
The uncertainty on the normalisation obtained in this manner is \percent{15} (\percent{30}) in the $b$-tagged ($b$-vetoed) sample, 
primarily due to uncertainties on the $b$-tagging efficiency and jet energy scale.

\paragraph{Estimation of the multi-jet background:}
\label{sec:estimation_llqcd}

The multi-jet background is estimated using the $ABCD$ method, by splitting the event sample into four regions
according to the charge product of the $e\mu$ pair
and the isolation requirements on the electron and muon. These requirements are summarised in
\tabref{tab:ABCD_QCD}.
\begin{table}
\centering
\begin{tabular}{ccc}
\hline\hline
Region & Charge correlation & Lepton isolation requirement\\
\hline
$A$ (Signal Region) & Opposite sign & isolated \\
$B$ & Same sign & isolated \\
$C$ & Opposite sign & anti-isolated \\
$D$ & Same sign & anti-isolated \\
\hline\hline
\end{tabular}
\caption{\label{tab:ABCD_QCD}Control regions for the estimation of the multi-jet background for the
\Htotautauemu{} and
\Htotautautolh{} samples: events are categorised according to the charge product of the two $\tau$ leptons and the lepton isolation requirement. 
In the \Htotautautolh{} channel isolation refers to the isolation of the electron or muon and in the \Htotautauemu~channel 
both the electron and muon are required to be isolated or anti-isolated, respectively.}
\end{table}

The systematic uncertainty of this method has been estimated by considering the stability of the ratio $r_{C/D}$
in regions where the isolation requirements are varied, or where only the muon is required to fail
the isolation. The resulting uncertainty on the normalisation is \percent{14} (\percent{23}) in the $b$-tagged ($b$-vetoed) sample.

Smaller backgrounds from \Wjets, \Ztoee, \Ztomumu, diboson, and single-top processes are
estimated from simulation.

\paragraph{Results:} 

\begin{figure}
  \centering
  \includegraphics[width=0.44\textwidth]{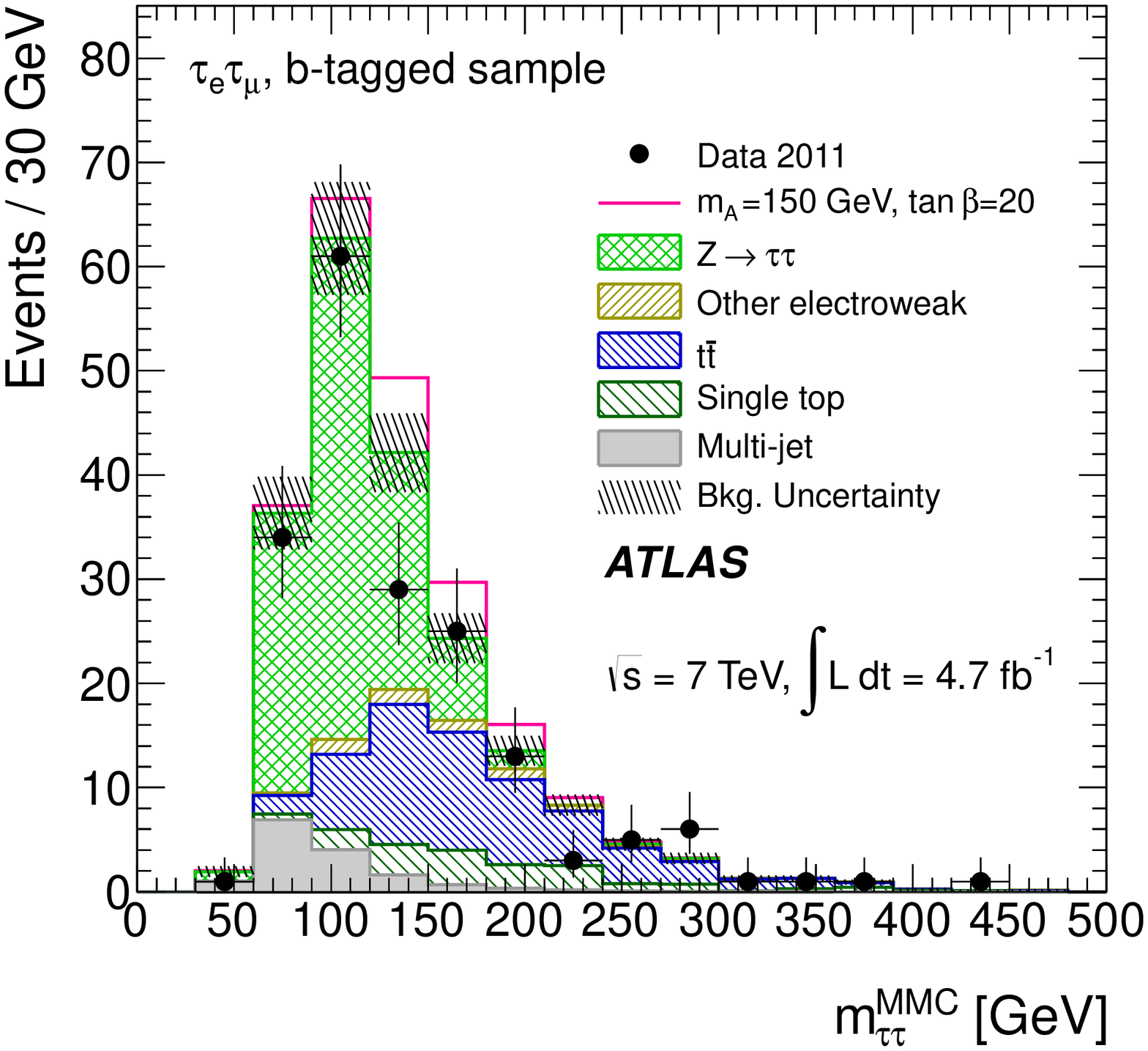}
  \includegraphics[width=0.44\textwidth]{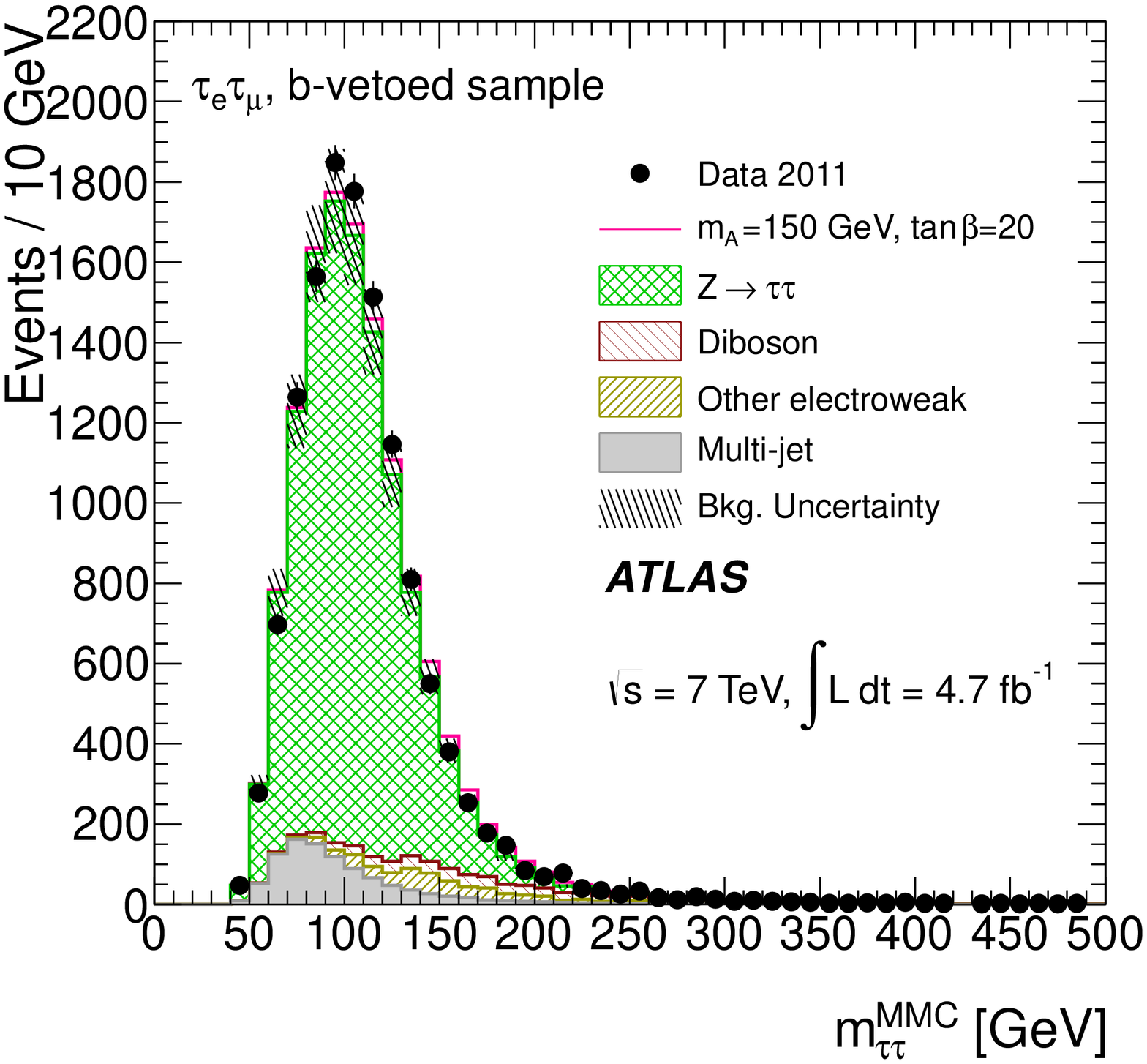}
  \caption{MMC mass distributions for the \Htotautauemu{} final state. The MMC mass, \mmcmass , is shown for the $b$-tagged (left-hand side) and $b$-vetoed samples (right-hand side). 
 The data are compared to the background expectation and an added hypothetical MSSM signal ($\MA=\unit[150]{\GEVcc}$ and $\tan\beta=20$).
 The background uncertainties include statistical and systematic uncertainties.
The contributions of the backgrounds \Ztoee , \Ztomumu{} and \Wjets{} are combined and labelled ``Other electroweak'';
in the case of the $b$-tagged samples the contributions of diboson production processes are included
as well. Background contributions from top quarks are included in ``Other electroweak'' for the
$b$-vetoed sample.
}
  \label{fig:leplep_finalmassplots}
\end{figure}

The number of observed $\telec\tmuon$ events in data, along with  predicted event yields from 
background processes, is shown in \tabref{tab:emu_summary_table}.
The observed event yield is compatible with the expected event yield from Standard Model processes
within the uncertainties. 
The MMC mass distributions for these events are shown in \figref{fig:leplep_finalmassplots}.

\begin{table}[htdp]
\centering
\renewcommand\arraystretch{1.2} 
\input{./leplep_yield.tex}

\caption{The number of events observed in data and the expected number of signal and background events of 
the \Htotautauemu~channel. Simulated event yields are normalised to the integrated luminosity of the data sample, \currentlumi . 
The predicted signal event yields correspond to a parameter choice of $\MA = \unit[150]{\GEVcc}$
and $\tanb = 20$ and include both the $b$-associated and the gluon-fusion production processes. 
\label{tab:emu_summary_table}
}
\end{table}
\renewcommand\arraystretch{1.}

%% file: leplep_yield.tex
\begin{tabular}{c r l r l}
\\
\hline \hline
		& \multicolumn{2}{c}{ $b$-tagged sample } &   \multicolumn{2}{c}{$b$-vetoed sample }	\\
\hline

\Ztotautau       & $109$    & $\pm 12$      & $11000$ & $\pm 1000$ \\
\Wjets	         &   $1.2$  & $\err{1.1}{0.9}$    &   $111$ & $\pm 23$ \\
\Ztoll	         &   $1.1$  & $\pm 0.8$           &   $196$ & $\err{22}{23}$ \\
\ttbar{}         &  $56$    & $\err{11}{9}$	  &   $150$ & $\err{60}{50}$ \\
Single top       &  $16$    & $\err{3}{4}$        &    $35$ & $\pm 5$ \\
Diboson          &   $3.9$  & $\pm 0.7$ 	  &   $470$ & $\pm50$ \\
Multi-jet        &  $15$    & $\pm 11$	          &   $980$ & $\pm230$  \\
\hline
Total 	         & $201$    & $\err{20}{19}$            & $13000$ & $\pm 1000$ \\ 
\hline
Signal $m_A=\unit[150]{GeV}, \tan\beta=20$  &  & & &\\
$b\bar{b}$(\Higgses$\rightarrow\tau\tau$)   &  $18$   & $\err{4}{5}$  & $270$ & $\err{40}{50}$ \\
$gg\rightarrow$\Higgses$\rightarrow\tau\tau$&   $2.3$ & $\pm  0.8$    & $143$ & $\err{23}{21}$\\
\hline
Data		& \multicolumn{2}{l}{$181$}  &   \multicolumn{2}{l}{$12947$} \\
\hline \hline
\end{tabular}

%% file: lephad.tex
\newpage
\subsection{The \Htotautautolh\ decay channel} \label{sec:lephad}

\paragraph{Signal topology and event selection:} \label{eq:selection_lephad}

Events in the \Htotautautolh\ channel are selected using a single-lepton trigger with transverse momentum 
thresholds of \unit[20]{\GEVc} or \unit[22]{\GEVc} for electrons, depending on the run period, and \unit[18]{\GEVc} for muons. 
Each event must contain one isolated
electron with $\pt>\unit[25]{\GEVc}$ or one isolated muon with $\pt>\unit[20]{\GEVc}$.
Events containing additional electrons or muons  
with transverse momenta greater than 
\unit[15]{\GEVc} or \unit[10]{\GEVc}, respectively, are rejected in order to obtain an orthogonal selection to 
those used in  the $\telec\tmuon$ and \mumu\ channels.
One \thad{} with a charge of opposite sign to the selected electron or muon is required. 
The \thad{} identification criterion in use 
is the one with medium efficiency as introduced in \secref{sec:objects}.
The transverse mass of the lepton and the missing transverse momentum, \mT, is required to be
less than \unit[30]{\GEVc},
to reduce contamination from \Wjets{} and \ttbar{} background processes.
Here the transverse mass is defined as:
\begin{eqnarray}
\mT =
\sqrt{2p_{\mathrm{T}}^{\mathrm{lep}}E_{\mathrm{T}}^{\mathrm{miss}}\left(1-\cos\Delta\phi\right)}~,
\end{eqnarray}
where $p_{\mathrm{T}}^{\mathrm{lep}}$ denotes the transverse
momentum of the electron or muon and $\Delta\phi$ 
the angle between $p_{\mathrm{T}}^{\mathrm{lep}}$ and \MET. 

After imposing these selection criteria, the resulting event sample is split 
into two categories depending on whether or not
the highest-$E_T$ jet with $\left|\eta\right|<2.5$ is identified as a \bjet. 
Events are included in the $b$-tagged sample 
if the highest-$E_T$ jet is identified as a \bjet\ and its $E_T$ is in the range of \unit[20]{\GEVc} to \unit[50]{\GEVc}.
Events are included in the $b$-vetoed sample if the highest-$E_T$ jet fails the \bjet\ identification 
criterion and the event has $\MET{} > \unit[20]{\GEVc}$.

\paragraph{Estimation of the $\boldsymbol{\Wjets{}}$ background:} \label{eq:W_normalization_lephad}
\Wjets{} events that pass the event selection criteria up to the \mT{} requirement consist primarily of events in
which the selected lepton originates from the $W$ decay and a jet is misidentified as a $\thad$.
To ensure a proper estimation of the jet-to-$\thad$ misidentification rate,
the $\Wjets$ background normalisation is corrected using
control regions with high purity in $\Wjets$ events  defined by requiring high transverse mass: $\unit[70]{\GEVc} < \mT < \unit[110]{\GEVc}$.
Separate control regions are used for the $\telec\thad$ and $\tmuon\thad$ samples, as the kinematic selections are different.
The correction factors derived from these control regions are ${f_W^{e} = 0.587 \pm 0.009}$ for the electron channel and 
${f_W^{\mu} = 0.541 \pm 0.008}$ for the muon channel, where the quoted uncertainties are statistical. 
The relative systematic uncertainty is estimated to be \percent{5} by varying the $\mT$ boundary definition of the control region. 
The correction factors have been derived separately for the $b$-tagged and $b$-vetoed samples; the numbers are in agreement between the two cases, but for the $b$-tagged sample the statistical uncertainty is \percent{17}. 
This statistical uncertainty is considered as an additional systematic uncertainty in the $b$-tagged sample analysis.

\paragraph{Estimation of the $\boldsymbol{\Ztotautau{}/e^{+}e^{-}/\mumu}$ background:}
The \Ztotautau{} background is estimated using the $\tau$-embedded {\Ztomumu} sample outlined in \secref{sec:embedding}.
The jet activity in the embedded events is independent of the $Z$ boson decay mode. 
Taking advantage of this feature, the embedding sample is also used to 
validate the simulated \Ztoee{} and \Ztomumu{} background samples for the correct \bjet\ fraction, which may affect
the background estimation after imposing the $b$-tag requirement.
Correction factors are derived by comparing $\tau$-embedded {\Ztomumu{}} events with simulated {\Ztotautau{}} events 
before and after the $b$-tagged sample selection. 
The correction factors are calculated to be $f_{Zb}^e=1.08\pm0.23$ and $f_{Zb}^\mu=1.11\pm 0.13$ for the 
electron and muon channels, respectively, where the quoted uncertainties are statistical.
The effect on these correction factors from the $t\bar{t}$ contribution in the control region is studied by removing the 
\unit[50]{\GEVc} maximum $E_T$ requirement on the \bjet . A \percent{7} systematic uncertainty is obtained. These factors are applied to the simulated \Ztoee{} and \Ztomumu{} background samples passing the $b$-tagged sample selection.

\paragraph{Estimation of the $\boldsymbol{\TTBAR{}}$ background:} \label{sec:ttbarCorrection_lephad}
The simulated \TTBAR{} samples are normalised from data using a top-enriched control region.
This control region is defined by applying the $\tlep\thad$ selection criteria
up to the requirement of a $e\thad$ or $\mu\thad$ pair in the event, with no requirement on the transverse mass.
The highest-$E_T$ jet in the event must be identified as a \bjet, with $E_T$ in the range \unit[$50$]{\GEVc} to \unit[$150$]{\GEVc}, 
and a second highest-$E_T$ jet must satisfy the same \bjet{} identification requirement. 
This results in a  control region with a purity of \TTBAR{} events over \percent{90} and negligible signal contribution.
The \ttbar{} correction factor is derived in a manner similar to that of the $\Wjets$ correction factor, and
a value of $f_{\ttbar} = 0.88$ $\pm\, 0.04$ (stat.) $\pm\, 0.14$ (syst.) is obtained with the systematic uncertainty due
primarily to the $b$-jet identification efficiency.

\paragraph{Estimation of the multi-jet background:} \label{sec:lephad_ABCD}
For the multi-jet background estimation, the $ABCD$ method is used by defining four regions
according to whether the charge of the $\tau$ jet and lepton
have opposite sign  or same sign, and whether the selected lepton passes or fails the isolation criteria.
These requirements are summarised in \tabref{tab:ABCD_QCD}.
In regions $C$ and $D$ the contribution from processes other than the  multi-jet background is negligible, while
in region $B$ there is a significant contribution from other backgrounds, in particular 
$Z/\gamma^{*}\, +$jets and \Wjets{}, which is subtracted from the data sample using estimates from simulation.
The systematic uncertainty on the predicted event yield is estimated by varying the definitions of the regions 
used, and by testing the stability of the $r_{C/D}$ ratio across the \mmcmass{} range.
The resulting uncertainty is \percent{7.5} in the \tmuon\thad{} channel and \percent{15} in the \telec\thad{} channel.

\begin{figure}
  \centering
  \includegraphics[width=0.44\textwidth]{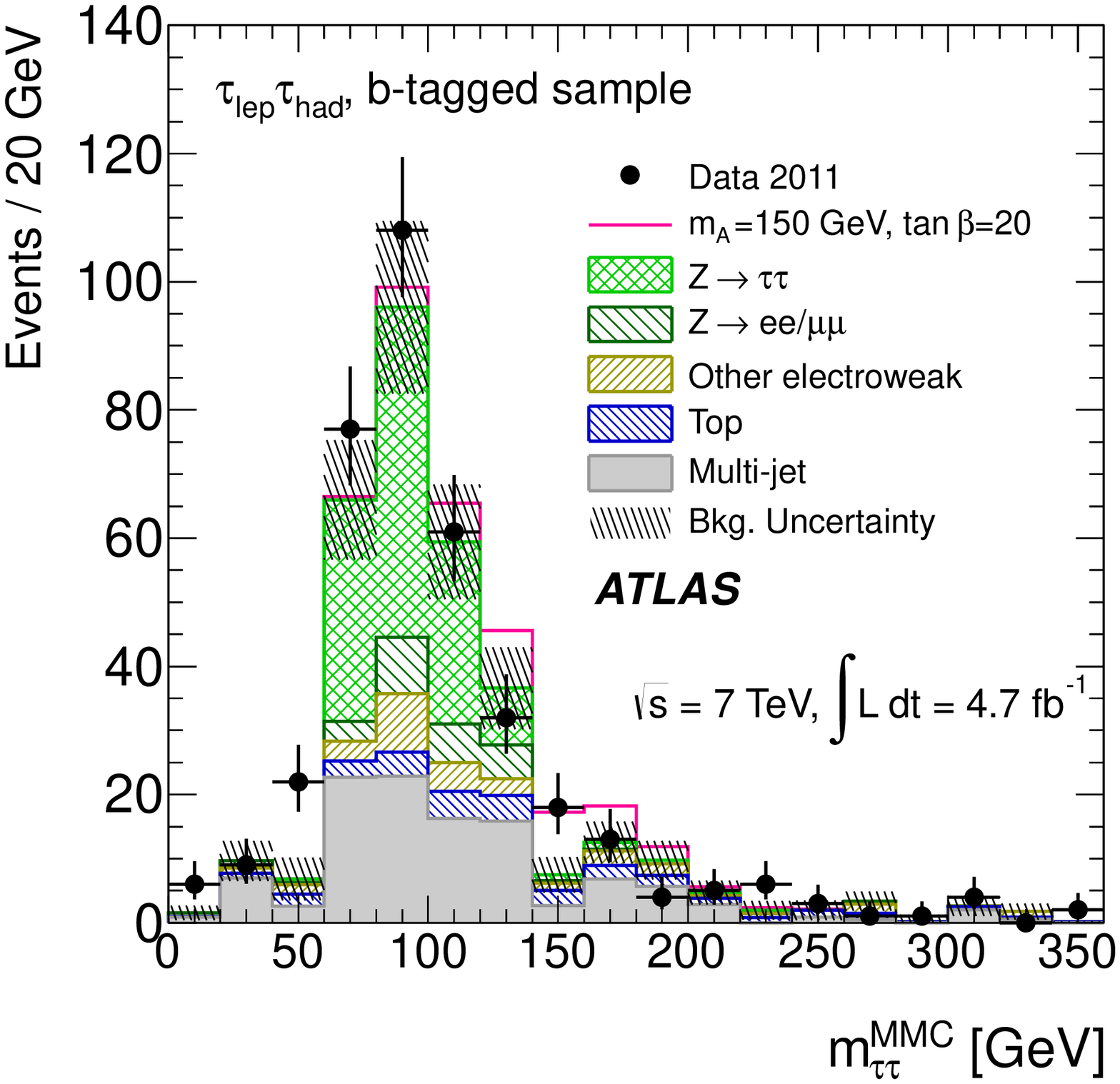}
  \includegraphics[width=0.44\textwidth]{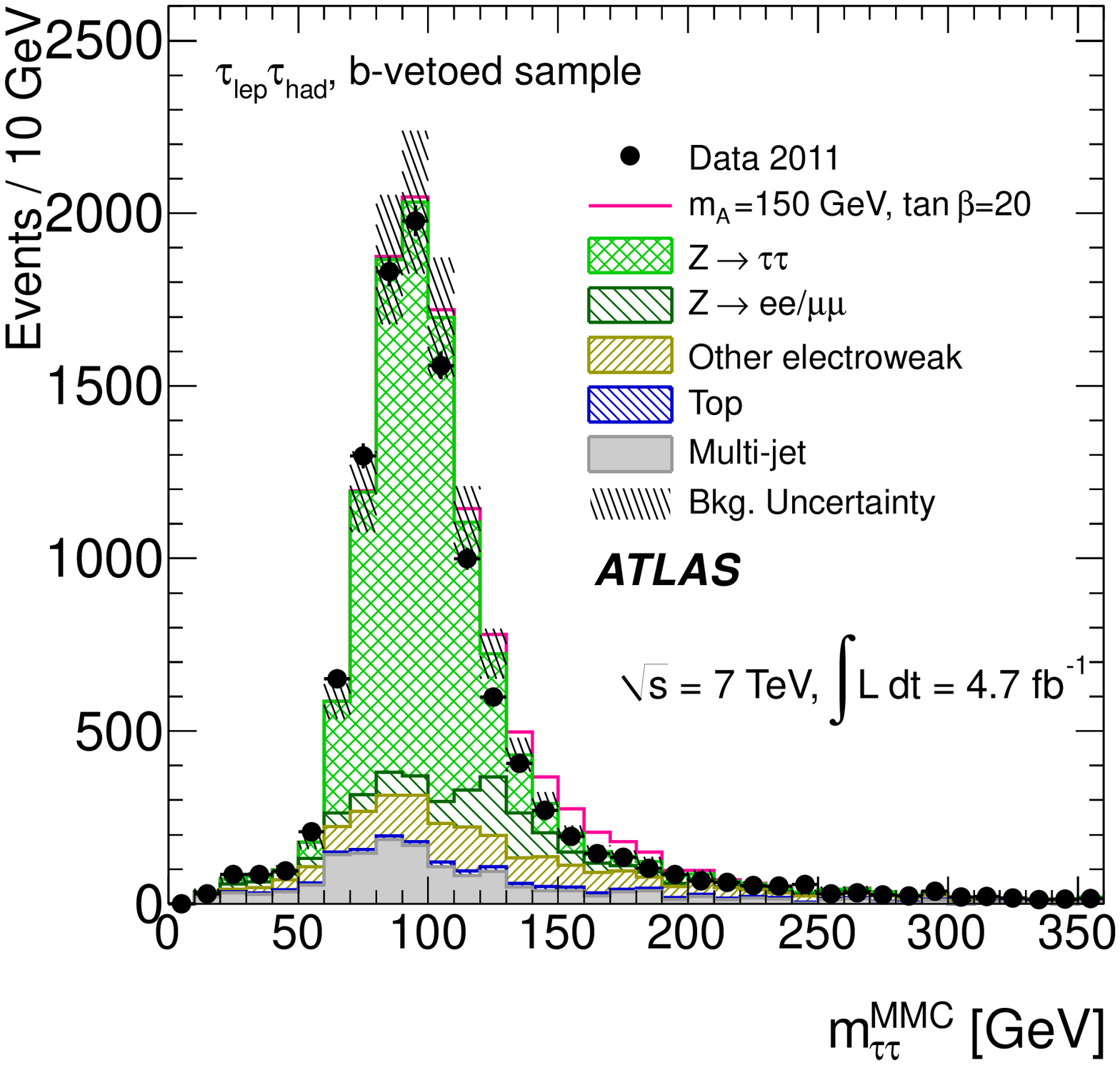}
  \caption{MMC mass distributions for the \Htotautautolh{} final state. The MMC mass,  \mmcmass, is shown 
for the $b$-tagged (left-hand side) and $b$-vetoed samples (right-hand side)  for the combined $\telec\thad$ and $\tmuon\thad$ samples. The data are compared to the background expectation and a
	hypothetical MSSM signal ($\MA=\unit[150]{\GEVcc}$ and $\tan\beta=20$). 
The background uncertainties include statistical and systematic uncertainties.
The contributions of the diboson and \Wjets{} background processes are combined and labelled ``Other electroweak''.
}
    \label{fig:lephad_finalmassplots}
\end{figure}

\paragraph{Results:} \label{sec:lephad_results}
The number of observed $\tlep\thad$ events in data, along with  predicted event yields from 
background processes, are shown in \tabref{tab:lephad_eventyield}.
The observed event yields are compatible with the expected yields from Standard Model processes
within the uncertainties. %The expected signal yield for $\MA=\unit[150]{\GEVcc}$ and ${\tan\beta=20}$ is shown for comparison.
The MMC mass distributions for these events, with $\tmuon\thad$ and $\telec\thad$ statistically combined,
are shown in \figref{fig:lephad_finalmassplots}.

\begin{table}[htdp]
\begin{center}
\centering
\renewcommand\arraystretch{1.2} %avoid overlap
\input{./lephad_yield.tex}

\caption{The number of events observed in data and the expected number of signal and background events 
for the \Htotautautolh~channel. 
Simulated event yields are normalised to the integrated luminosity of the data sample, \currentlumi.
The predicted signal event yields correspond to the parameter choice $\MA = \unit[150]{\GEVcc}$ and $\tanb = 20$ 
and include both the $b$-associated and the gluon-fusion production processes.
\label{tab:lephad_eventyield}
}
\end{center}
\end{table}

\renewcommand\arraystretch{1.}

%% file: lephad_yield.tex
\begin{tabular}{ c r l r l}
\\ \hline \hline

			& \multicolumn{4}{c}{ Muon Channel ($\tmuon\thad$) } \\   
			& \multicolumn{2}{c}{ $b$-tagged sample } & \multicolumn{2}{c}{$b$-vetoed sample } \\ \hline \hline
\Ztotautau		&  86    &  $\pm 15 $ 	                & 4800 & $\pm 700$ \\
\Wjets			&  19    &  $ \err{6}{8} $ 	        &  780 & $\err{100}{140} $ \\
\Ztoll			&   8    &  $ \err{5}{4} $ 	        &  350 & $ \err{100}{90} $ \\
Top			&  14.5  &  $ \err{3.5}{ 2.7 } $ 	&  105 & $ \err{20}{21} $ \\
Diboson		        &   0.8  &  $ \pm 0.4 $ 	        &   38 & $ \err{6}{5} $ \\
Multi-Jet	        &  51    &  $ \pm 11 $	                &  580 & $ \err{140}{130} $ \\ \hline
Total		        & 180    &  $ \pm 20$  	                & 6600 & $ \pm 800$ \\ \hline
Signal $m_A=\unit[150]{GeV}, \tan\beta=20$      &   &  & &  \\
$b\bar{b}$(\Higgses$\rightarrow\tau\tau$)  	&	20   &  $\err{5}{6} $  & 174 & $\err{27}{35} $ \\
$gg\rightarrow$\Higgses$\rightarrow\tau\tau$	&	1.2  &  $\pm 0.6    $  & 115 & $\pm 16 $       \\ \hline
Data			&	\multicolumn{2}{l}{202	}		       & \multicolumn{2}{l}{6424	} \\ \hline \hline

\\ \hline\hline

			& \multicolumn{4}{c}{ Electron Channel ($\telec\thad$) } \\   
			& \multicolumn{2}{c}{$b$-tagged sample } & \multicolumn{2}{c}{$b$-vetoed sample } \\ \hline \hline
\Ztotautau		&   $42$   &  $\pm 20$	         &   $2700$ &  $ \pm 500 $ \\
\Wjets			&   $18$   &  $\err{9}{12}$	 &    $740$ &  $ \err{110}{160} $ \\
\Ztoll			&   $19$   &  $\pm 10$           &    $700$ &  $ \err{350}{270} $ \\
Top			&   $15.1$ &  $\pm 3.0$          &    $106$ &  $\err{20}{21} $ \\
Diboson			&    $1.0$ &  $\err{0.4}{0.5}$   &     $29$ &  $\err{5}{4} $ \\
Multi-Jet		&   $60$   &  $\pm 15$ 	         &    $920$ &  $\err{230}{240} $	\\ \hline
Total			&  $154$   &  $\pm 30$	         &   $5200$ &  $\pm 600$ \\ \hline
Signal $m_A=\unit[150]{GeV}, \tan\beta=20$ & & &  & \\
$b\bar{b}$(\Higgses$\rightarrow\tau\tau$)	& $15$   &  $\err{3}{5} $    & $138$ & $\err{22}{29} $ \\
$gg\rightarrow$\Higgses$\rightarrow\tau\tau$	& $1.2$  &  $\err{0.6}{0.4}$ &  $99$ & $\err {15}{14} $ \\ \hline
Data						&  \multicolumn{2}{l}{$175$} &  \multicolumn{2}{l}{$5034$}	\\ \hline \hline

\end{tabular}

%% file: hadhad.tex
\def\Selectedbtag{\ensuremath{27}}
\def\Selectednobtag{\ensuremath{1223}}
\def\Bkgdbtag{\ensuremath{25.4\pm 5.2}}
\def\Bkgdnobtag{\ensuremath{1232 +72/-63}}
\def\BkgdSysbtag{\ensuremath{25.4\pm 4.8\,{}^{+2.2}_{-1.9}}}
\def\BkgdSysnobtag{\ensuremath{1232\pm 23\,{}^{+68}_{-59}}}

\subsection{The \Htotautautohh\ decay channel} \label{sec:hadhad}

\paragraph{Signal topology and event selection:}

Events in this channel are selected by a di-\thad{} trigger with transverse momentum 
thresholds of \unit[29]{\GEVc} and \unit[20]{\GEVc} for the two \thad{} candidates.
Events containing identified electrons or muons with transverse momenta above 
\unit[15]{\GEVc} or \unit[10]{\GEVc}, respectively, are vetoed. 
These vetoes suppress background events and ensure  that the channels are statistically independent.
Two $\thad$ candidates with opposite-sign charges are required, 
one passing the tight \thad{} identification requirements 
and the second passing the medium criteria. 
These two leading $\thad$ candidates are required to match the 
reconstructed $\thad$ trigger objects each within a cone with radius parameter $\Delta R < 0.2$.
These two $\thad$ candidates are required to have transverse 
momenta above \unit[45]{\GEVc} and \unit[30]{\GEVc}, respectively.
These values are chosen such that the plateau of the 
trigger turn-on curve is reached and the electroweak and multi-jet backgrounds are suppressed effectively.
The missing transverse momentum is required to be above \unit[25]{\GEVc} to account for the presence of neutrinos 
originating from the $\tau$ decays and to suppress multi-jet background.

The selected events are split into a $b$-tagged sample and a $b$-vetoed sample to exploit the two dominant production mechanisms for neutral Higgs bosons in the MSSM.
Events in which the leading jet is identified as a \bjet\ are included in the $b$-tagged sample. 
The transverse momentum of this jet is restricted to the range of \unit[$20$]{\GEVc} to \unit[$50$]{\GEVc}
to reduce the \ttbar{} background.
Events without jets, or in which the leading jet is not identified as a \bjet , 
are included in the $b$-vetoed sample.
Due to the higher background levels in this sample, the threshold 
on the transverse momentum of the leading $\thad$ candidate 
is raised to \unit[60]{\GEVc}.  

\paragraph{Identification efficiency and misidentification corrections for hadronic $\boldsymbol{\tau}$ decays:}
\label{triggerEfficiency}

The \thad{} identification efficiencies, 
the \thad{} trigger efficiencies and the corresponding misidentification
probabilities are corrected for differences observed between data and simulation.
For the di-\thad{} trigger it is assumed 
that these identification and misidentification efficiencies
can be factorised into 
the efficiencies of the corresponding 
single-\thad{} triggers with 
appropriate transverse momentum requirements. 
This factorisation is validated using a simulated event sample.
The single-\thad{} trigger efficiency for real hadronically decaying \atau{} leptons 
with respect to the offline \thad{} selection was measured using a tag-and-probe analysis with 
$Z\rightarrow\tmuon\thad$ data. A correction factor for the simulation was derived 
as a function of the transverse momentum of each of the two $\thad$ candidates, and each event was weighted by the product of these factors.
The probability to misidentify a jet as a $\thad$
is extracted for both the trigger and the \thad~identification algorithm by analysing jets in a high-purity \Wtomunu\ sample.  
A correction factor derived on the basis of these probabilities is applied to the simulation when a jet is misidentified as a $\thad$.
The statistical and systematic uncertainty of these correction factors leads to an uncertainty of \percent{21} on the \Wjets{} background and \percent{4} to \percent{6} on the $\ttbar$ background.

\paragraph{Estimation of the $\boldsymbol{\Ztotautau{}}$ and $\boldsymbol{\Wtotaunu{}}$ backgrounds:}

The estimates of the \Ztotautau{} and \Wtotaunu{} backgrounds are taken from simulation
and are validated using $\tau$-embedded \Ztomumu{} and \Wtomunu{} samples. The $\tau$-embedded \Ztomumu{} data are 
used to validate the simulation, rather than to provide the main estimate for the \Ztotautau{} background, 
in the \Htotautautohh{} channel. The di-\thad{} trigger is not modelled in the $\tau$-embedded \Ztotautau{} data, 
making it difficult to apply the embedding technique. 
Correction factors for the efficiency of the \bjet{} identification requirement on the leading jet
are derived in a way equivalent to that described in \secref{sec:lephad}. 
For the \Zjets{} background a factor of $f_{Zb} = 1.24\pm 0.34$ (stat.) is derived by comparing the 
simulated and embedded \Ztotautau{} samples.
With the embedded \Wtotaunu{} sample, 
no such correction factor may be derived in this way as the contamination from 
\ttbar{} events is quite significant once the \bjet{} identification requirements are applied.
Instead, the procedure in \secref{sec:lephad} is applied with the selection of this channel. 
A correction factor of $f_{Wb} = 1.00\pm 0.31$ (stat.) is derived for \Wtotaunu{} events.
The correction factors for the $b$-vetoed sample are found to be close to unity and uncertainties are negligible, 
hence no correction is applied.

\paragraph{Estimation of the multi-jet background:}
\label{sec:hadhad_QCD}

The multi-jet background is estimated using the $ABCD$ method, by splitting the event sample into four regions 
based on the charge product of the two leading $\thad$ candidates
and whether the nominal \thad{} identification requirements of these two $\thad$ candidates are met.
These variables can be assumed to be uncorrelated for multi-jet events. \Tabref{tab:ABCDdefinition} illustrates the definition of the four regions.

\begin{table}
\centering
\begin{tabular}{ccc}
\hline\hline
Region & Charge correlation & Hadronic $\tau$ decay identification requirement\\
\hline
$A$ (Signal Region) & Opposite sign & Pass  \\
$B$ & Same sign & Pass \\
$C$ & Opposite sign & Fail  \\
$D$ & Same sign & Fail  \\
\hline\hline
\end{tabular}
\caption{\label{tab:ABCDdefinition}Control regions for the estimation of the multi-jet background for the
\Htotautautohh{} selection: Events are categorised according to the product of the electric charges of the two $\thad$ candidates and the $\thad$ identification. 
``Pass'' refers to one $\thad$ candidate passing the tight and the other $\thad$ candidate passing the medium identification selection. 
``Fail'' refers to all events in which the two $\thad$ pass the loose identification selection but do not satisfy the selection of the ``Pass'' category.
}
\end{table}

The shape of the MMC mass distribution for the multi-jet background in the signal region
is taken from region $C$ for the $b$-vetoed sample and from region $B$ for the $b$-tagged sample. 
Contributions from electroweak backgrounds are subtracted from each control region using simulation.
The uncertainties on these backgrounds lead to uncertainties on the multi-jet estimate 
of \percent{7} for the $b$-tagged sample and \percent{5} for the $b$-vetoed sample.

\paragraph{Results:}

\begin{figure}
  \centering
 \subfigure{  \includegraphics[width=0.44\textwidth]{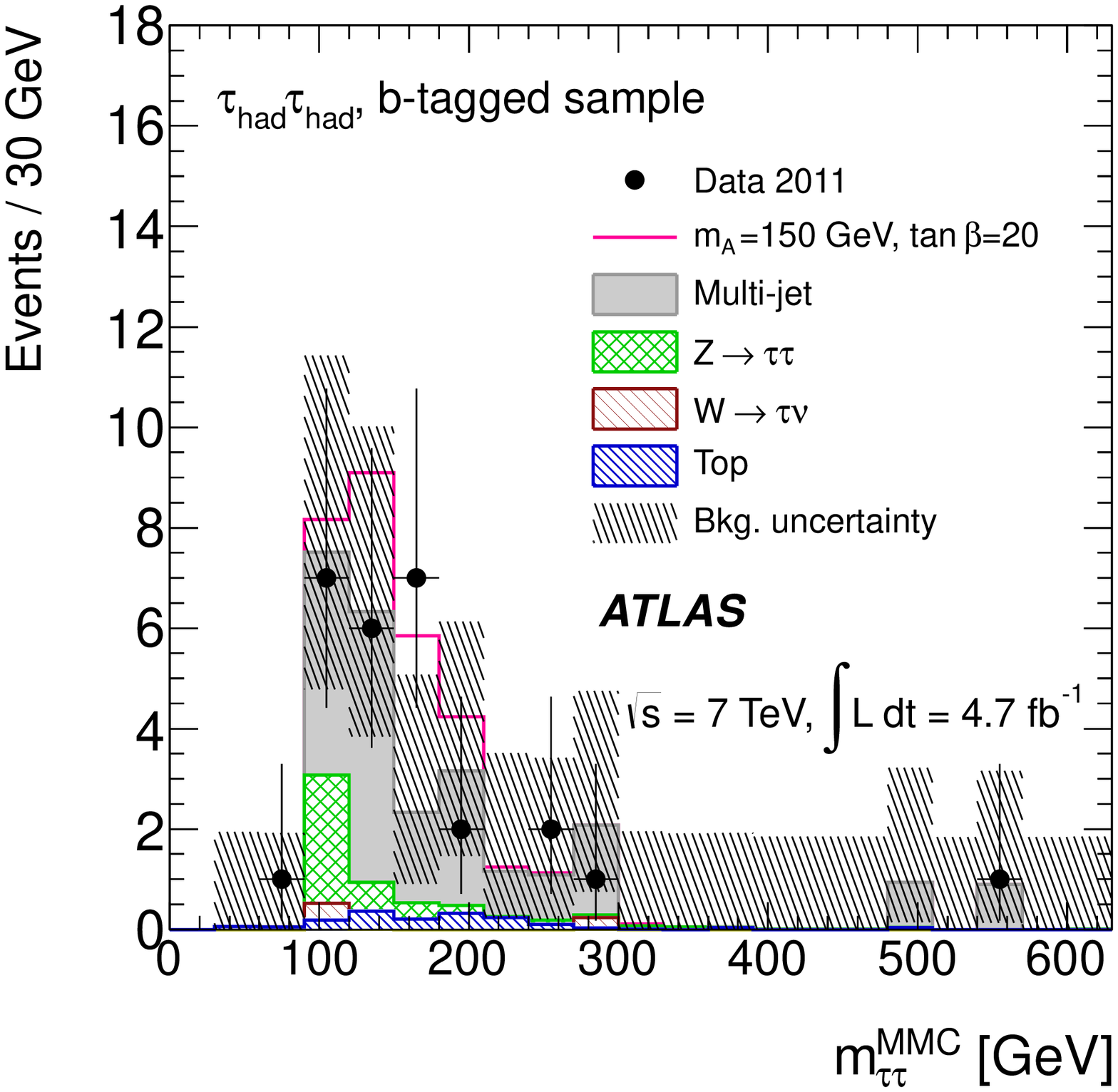}  }
  \subfigure{ \includegraphics[width=0.44\textwidth]{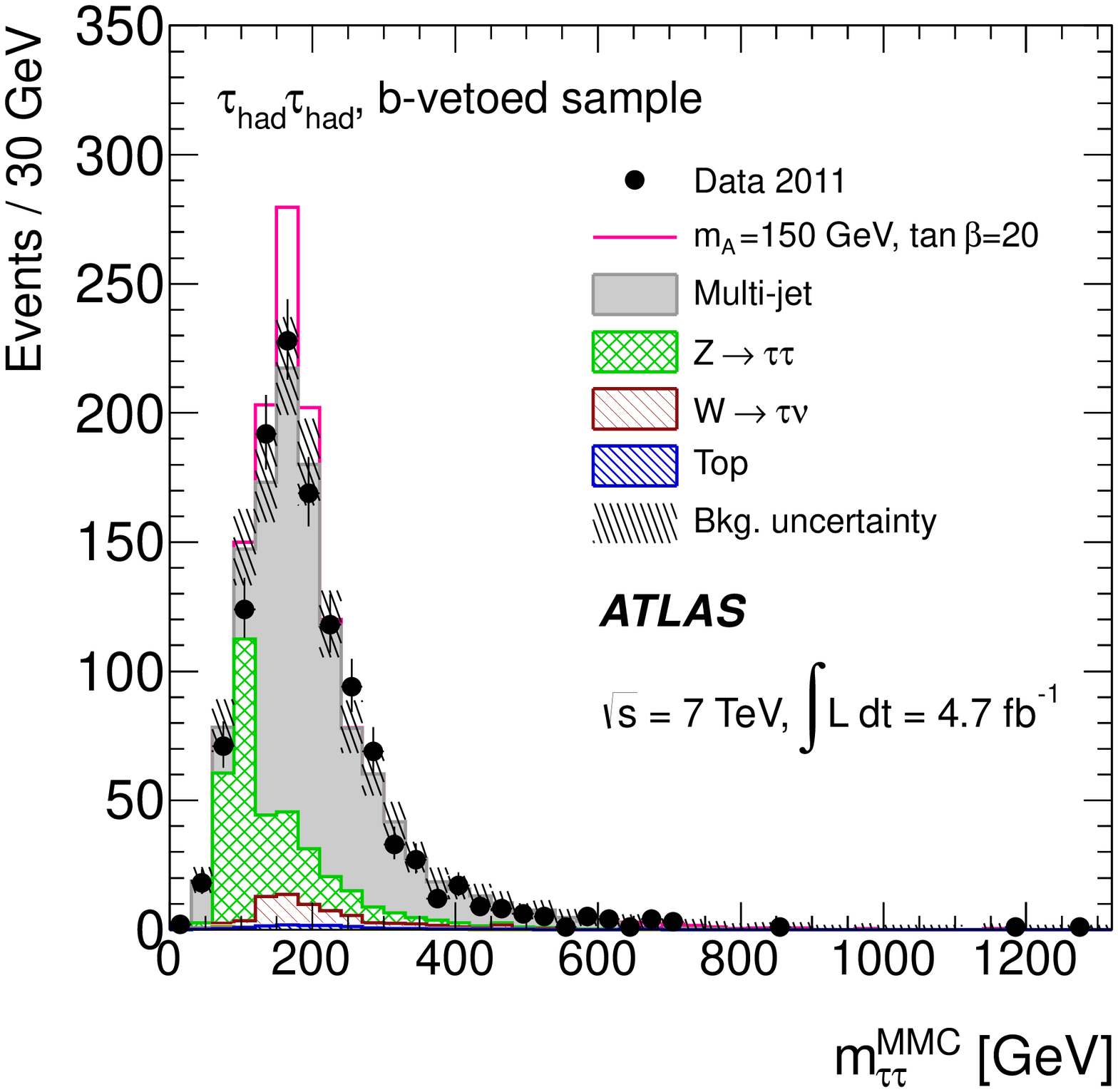}  } 
\caption{
MMC mass distributions for the \Htotautautohh{} final state. The MMC mass , \mmcmass , is shown for the $b$-tagged (left-hand side) and $b$-vetoed
    samples (right-hand side). The data are compared to the background expectation and an added hypothetical MSSM signal ($\MA=\unit[150]{\GEVcc}$ and $\tan\beta=20$). The background uncertainties include statistical and systematic uncertainties.
	} \label{fig:hadhad_finalmassplots}
\end{figure}

The number of observed $\thad\thad$ events in data, along with  predicted event yields from 
background processes, is shown in \tabref{tab:eventyields_hadhad}.
The observed event yields are compatible with the expected yields from Standard Model processes
within the uncertainties. %The expected signal yield for $\MA=\unit[150]{\GEVcc}$ and ${\tan\beta=20}$ is shown for comparison.
The MMC mass distributions for these events are shown in \figref{fig:hadhad_finalmassplots}.

\begin{table}[h]
\centering
\renewcommand\arraystretch{1.2}
\input{./hadhad_yield.tex}
\caption{The observed number of events in data and the expected number of signal and background events for the 
\Htotautautohh~channel. Simulated event yields are normalised to the total integrated luminosity, \currentlumi.
The data are compared to the background expectation and an added hypothetical MSSM signal 
($\MA = \unit[150]{\GEVcc}$ and $\tanb = 20$). 
Because of the subtraction of the \Ztotautau{} background in the control regions used in the 
multi-jet background estimation the uncertainties of the \Ztotautau{} and the 
multi-jet backgrounds are anti-correlated.
}
\label{tab:eventyields_hadhad}
\end{table}

\renewcommand\arraystretch{1.}

%% file: hadhad_yield.tex
\begin{tabular}{c r l r l}
\hline\hline
 & \multicolumn{2}{c}{$b$-tagged sample} & \multicolumn{2}{c}{$b$-vetoed sample}  \\
\hline\hline
Multi-jet         &  $19$    &  $\pm 5$            & $870$    &  $\pm 50$       \\
\Ztotautau        &  $4.0$   &  $\pm 3.0$          & $300$    &  $\err{80}{70}$ \\
\Wjets            &  $0.5$   &  $\err{0.5}{0.4}$   &  $50$    &  $\pm 20$       \\
Top               &  $1.7$   &  $\pm 0.6$          &  $11.2$  &  $\pm 2.2$      \\
Diboson           &  $0.01$  &  $\pm 0.04$         & $4.9$    &  $\pm 1.0$      \\
\hline
Total             & $25$     & $\pm 5$             & $1200$   & $\err{80}{70}$  \\
\hline
Signal $m_A=\unit[150]{GeV}, \tan\beta=20$    &  & &  & \\
$b\bar{b}$(\Higgses$\rightarrow\tau\tau$)     &  $7.7$  &   $\pm 3.4$  & $73$ &  $\pm 21 $  \\
$gg\rightarrow$\Higgses$\rightarrow\tau\tau$  &  $0.5$  &   $\pm 0.2$  & $47$ &  $\pm 11 $  \\
\hline
Data & \multicolumn{2}{l}{$27$} & \multicolumn{2}{l}{$1223$}  \\
\hline\hline
\end{tabular}

%% file: systematics.tex
\section{Systematic uncertainties} \label{sec:systematics}

\paragraph{Data-driven background estimation:}

Where possible, event yields and mass distributions for the background are estimated using
control samples in data. The specific techniques and their associated uncertainties have been
presented in the relevant sections.
The effect of these uncertainties on the predicted background event yield is less than \percent{5} for the \mumu{}
channels and less than \percent{15} for the \tautau{} channels, and is usually small compared to the
systematic uncertainties from the simulated samples.

\paragraph{Cross-section for signal and background samples:}

The uncertainties on the signal cross-sections are estimated to be \percent{10} to \percent{20},
depending on the values of $\MA$ and \tanb{},
for both gluon-fusion and $b$-associated Higgs boson production~\cite{LHCHiggsCrossSectionWorkingGroup:2011ti}.
An uncertainty of \percent{5} is assumed on the cross-sections for $W$ and $Z$ boson background production 
\cite{WZcrossSections1,WZcrossSections2}.
Uncertainties due to the parton distribution functions and the renormalisation and 
factorisation scales are included in these estimates.

\paragraph{Acceptance modelling for simulated samples:}

The uncertainty on the acceptance from the parameters used in the event generation 
of signal and background samples is also considered.
This is done by evaluating the change in kinematic acceptance after varying the relevant scale parameters, 
parton distribution function choices, and if applicable, conditions for the matching of the partons 
used in the fixed order calculation and the parton shower. Furthermore, the effects 
of different tunes of the underlying event activity are considered.
The resulting uncertainties are typically \percent{2} to \percent{20}, depending on the sample and the channel considered.

\paragraph{Electron and muon identification and trigger:}
The uncertainties on electron or muon
trigger and identification efficiencies are determined from data using samples of $W$ and $Z$
decays \cite{MuonEfficiencyConfNote,egammapaper}. The trigger efficiency uncertainties
are below about \percent{1}. The identification efficiency uncertainties are between \percent{3} and \percent{6}
for electrons and below \percent{1.8} for muons.
The total effect of these uncertainties on the event yield is no greater than \percent{4} in any channel.

\paragraph{Hadronic $\boldsymbol{\tau}$ identification and trigger:}

The uncertainties related to hadronic $\tau$ trigger and identification efficiencies
are also studied with data \cite{ATLASTauIDNew}. The di-$\thad$ trigger efficiency
used in the $\thad\thad$ channel has an uncertainty of \percent{2} to \percent{7}.
The identification efficiency uncertainty for hadronic $\tau$ decays is about \percent{4}
for reconstructed $\thad$  $\pt$ above \unit[22]{\GEVc} and \percent{8} otherwise.
These uncertainties are most important in the $\thad\thad$ channel,
where the effect on the estimated signal yield reaches \percent{11}.

\paragraph{$\boldsymbol{b}$-jet identification:}
The \bjet{} identification efficiency 
and the misidentification probabilities for jets other than \bjets{}
have been measured in data \cite{BtaggingEfficiency,BtaggingMistag}.
The associated uncertainties are treated
separately for all jet flavours and depend on jet $E_T$ and~$\eta$.
Typical uncertainty values are around \percent{5} for \bjets{} and between
\percent{20} and \percent{30} for other jets, leading to a total uncertainty close to
\percent{5} on the event yield in the $b$-tagged channels.

\paragraph{Energy scale and resolution:}
The uncertainty on the acceptance due to the energy measurement uncertainty in the calorimeter is considered for each 
identified object corresponding to the clusters in the calorimeters.
For the clusters identified as 
electrons, typically a \percent{1} (\percent{3}) energy scale uncertainty is assigned for the barrel 
(end-cap) region~\cite{egammapaper}. 
The energy scale uncertainties for clusters identified as 
hadronic $\tau$ decays and jets are treated as being fully correlated, and are typically around
\percent{3}~\cite{ATLASTauIDNew,ATLASJETEnergyScale}.
The uncertainty in the muon energy scale is below \percent{1}. 
The acceptance uncertainty due to the jet energy resolution, which affects
the $\pt$ thresholds used to define the 
$b$-tagged and $b$-vetoed samples, is typically less than \percent{1}.
The systematic uncertainty due to the energy scales of electrons, muons,
hadronic $\tau$ decays and jets is propagated to the \MET\ vector. 
Additional uncertainties due to different pile-up conditions in data and simulation
are also considered. 

The uncertainty on the acceptance due to the energy scale and resolution variations reaches up to \percent{37} for signal in the
\thad\thad{}  $b$-tagged channel, but is usually less than \percent{10} for channels with fewer \thad.

\paragraph{Luminosity:}
The simulated sample event yield is normalised to the integrated luminosity of the data, which is
measured~\cite{lumi1,lumi2} to be \unit[4.7]{fb$^{-1}$} and \unit[4.8]{fb$^{-1}$} for the $\tautau$
and $\mumu$ channels, respectively,
and has an uncertainty of \percent{3.9}. This is applicable to all signal and background processes
which are not normalised using data-driven methods.

%% file: statistics.tex
\section{Statistical analysis} \label{sec:statistics}

The statistical analysis of the data employs a binned likelihood function.
Each one of the $\mu\mu$, $\telec\tmuon$, $\telec\thad$, $\tmuon\thad$ and $\thad\thad$ final states  is split into a $b$-tagged and a $b$-vetoed sample.
The likelihood in each category is a product over bins in the distributions of the MMC mass in the signal and control regions.

The expected number of events for signal ($s_j$) and background ($b_j$), as well as the observed number of events ($N_j$) 
in each bin of the mass distributions, enter in the definition of the likelihood function $\mathcal{L}(\mu,\boldsymbol{\theta})$.
A ``signal strength'' parameter ($\mu$) scales the expected signal in each bin. The value $\mu=0$
corresponds to the background-only hypothesis, while $\mu=1$ corresponds to the
signal-plus-background hypothesis with all Higgs bosons having the masses and cross-sections
specified by the point considered in the $\MA$--$\tb$ plane for the MSSM exclusion limit.
Signal and background predictions depend on systematic uncertainties that are
parameterised by nuisance parameters, $\boldsymbol{\theta}$, which in turn are
constrained using Gaussian functions, $\mathcal{F}_{\text{G}}$, so that
\begin{equation}
\mathcal{L}\left(\mu, \boldsymbol{\theta}\right) = 
\prod\limits_{\substack{
   j\textrm{ = bin and}\\
   \textrm{category}}
  }
\mathcal{F}_{\text{P}}\left(N_j\right.\left|\,\mu\cdot s_j + b_j\right)
\prod_{\theta_i} 
\mathcal{F}_{\text{G}}\left(\theta_i\right.\left|\,0,1\right),
\end{equation}
where  $\mathcal{F}_{\text{P}}\left(N_j\right.\left|\,\mu\cdot s_j + b_j\right)$ denotes the Poisson distribution 
with mean $\mu\cdot s_j + b_j$ for variable $N_j$.
The correlations of the systematic uncertainties across categories are taken into account. The expected signal and background event counts in each bin are functions of
$\boldsymbol{\theta}$. The parameterisation is chosen such that the rates in each
channel are log-normally distributed for a normally distributed $\boldsymbol{\theta}$.

To calculate the upper limit on $\mu$ for a given signal hypothesis, the compatibility of the observed or expected dataset with the signal-plus-background prediction is checked following the modified frequentist method known as CL$_{s}$~\cite{CLs_2002}. 
The test statistic $\tilde{q}_\mu$, used in the upper limit derivation, is defined as 
\begin{equation}
\tilde{q}_\mu =\begin{cases}
    -2\ln\left(\frac{\mathcal{L}(\mu,\hat{\boldsymbol{\theta}}_\mu)}{\mathcal{L}(0,\hat{\boldsymbol{\theta}}_{0})}\right) & \quad\text{if } \hat{\mu}<0,\\
    -2\ln\left(\frac{\mathcal{L}(\mu,\hat{\boldsymbol{\theta}}_\mu)}{\mathcal{L}(\hat{\mu},\hat{\boldsymbol{\theta}})}\right) & \quad\text{if } 0\leq \hat{\mu}\leq \mu ,\\
    0 & \quad\text{if } \hat{\mu} > \mu ,\\
\end{cases} \label{eq:qtilde}
\end{equation}
where $\hat{\mu}$ and $\hat{\boldsymbol{\theta}}$ refer to the global maximum of
the likelihood and
$\hat{\boldsymbol{\theta}}_\mu$ corresponds to the conditional maximum
likelihood for a given $\mu$.
The asymptotic approximation~\cite{CCGV} is used to evaluate the probability density functions rather than performing pseudo-experiments; the procedure has been validated using ensemble tests.

The significance of an excess in data is quantified with the local $p_{0}$-value, the probability that
the background processes can produce a fluctuation greater than or equal to the excess observed in data. 
The test statistic $q_{0}$, which is used in the local $p_{0}$-value calculation, is defined as:
\begin{equation}
q_0 =\begin{cases}
    -2\ln\left(\frac{\mathcal{L}(0,\hat{\boldsymbol{\theta}}_0)}{\mathcal{L}(\hat{\mu},\hat{\boldsymbol{\theta}})}\right) & \quad\text{if } \hat{\mu}\geq 0 ,\\
    0 & \quad\text{if } \hat{\mu} < 0 ,\\
\end{cases}
\end{equation}
where the notation is the same as in Equation \ref{eq:qtilde}. 
The equivalent formulation in terms of the number of
standard deviations ($\sigma$), $Z_{0}$, is referred to as the local significance and is defined as:
\begin{equation}
Z_0 = \Phi^{-1}(1-p_{0}),
\end{equation}
where $\Phi^{-1}$ is the inverse of the cumulative distribution of the Gaussian distribution.
The local $p_{0}$-value is estimated using the asymptotic
approximation for the $q_{0}$ distribution~\cite{CCGV}.

%% file: results.tex
\section{Results} \label{sec:results}

\begin{figure}
  \centering
  \includegraphics[width=.45\textwidth]{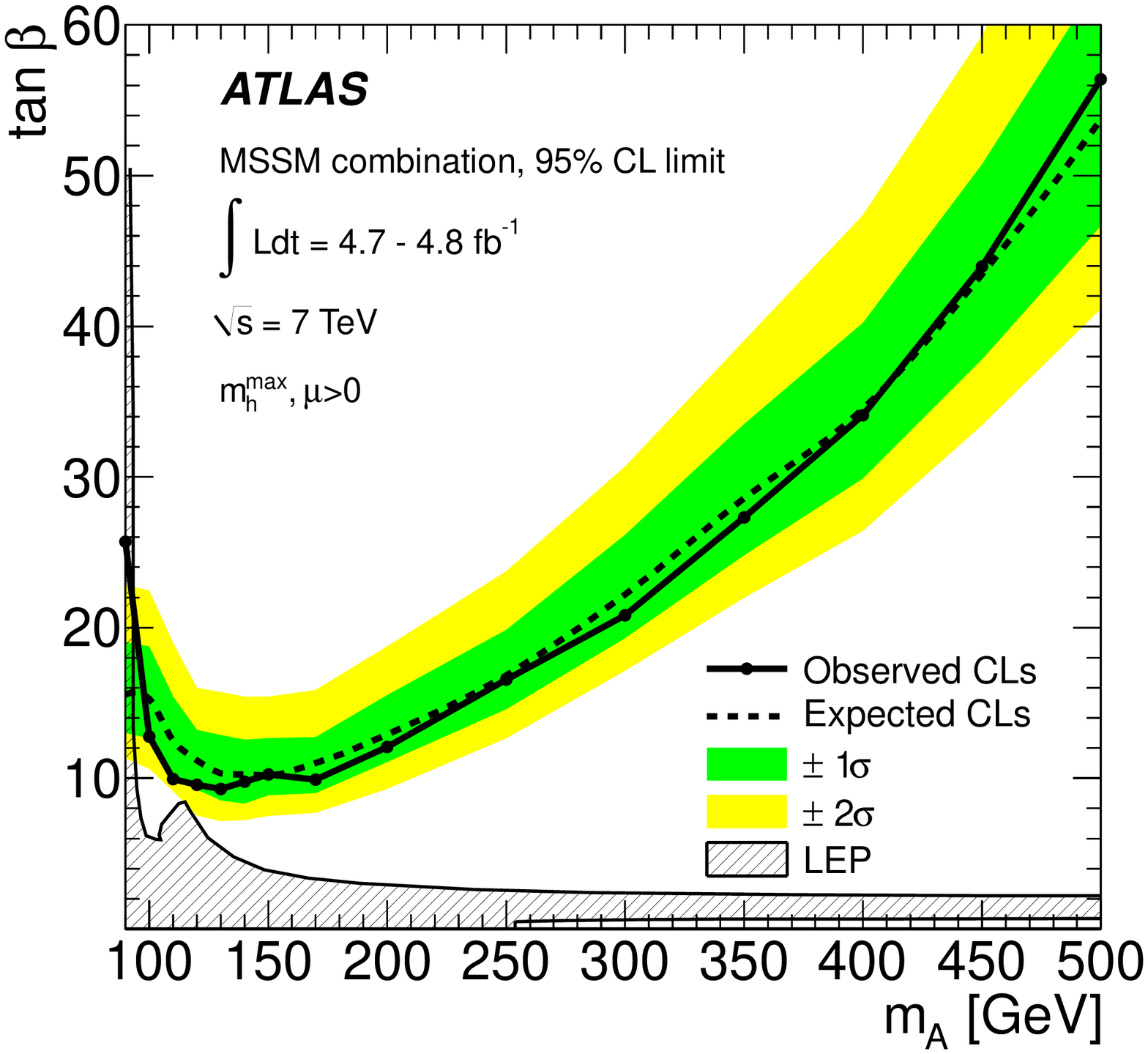}
  \includegraphics[width=.45\textwidth]{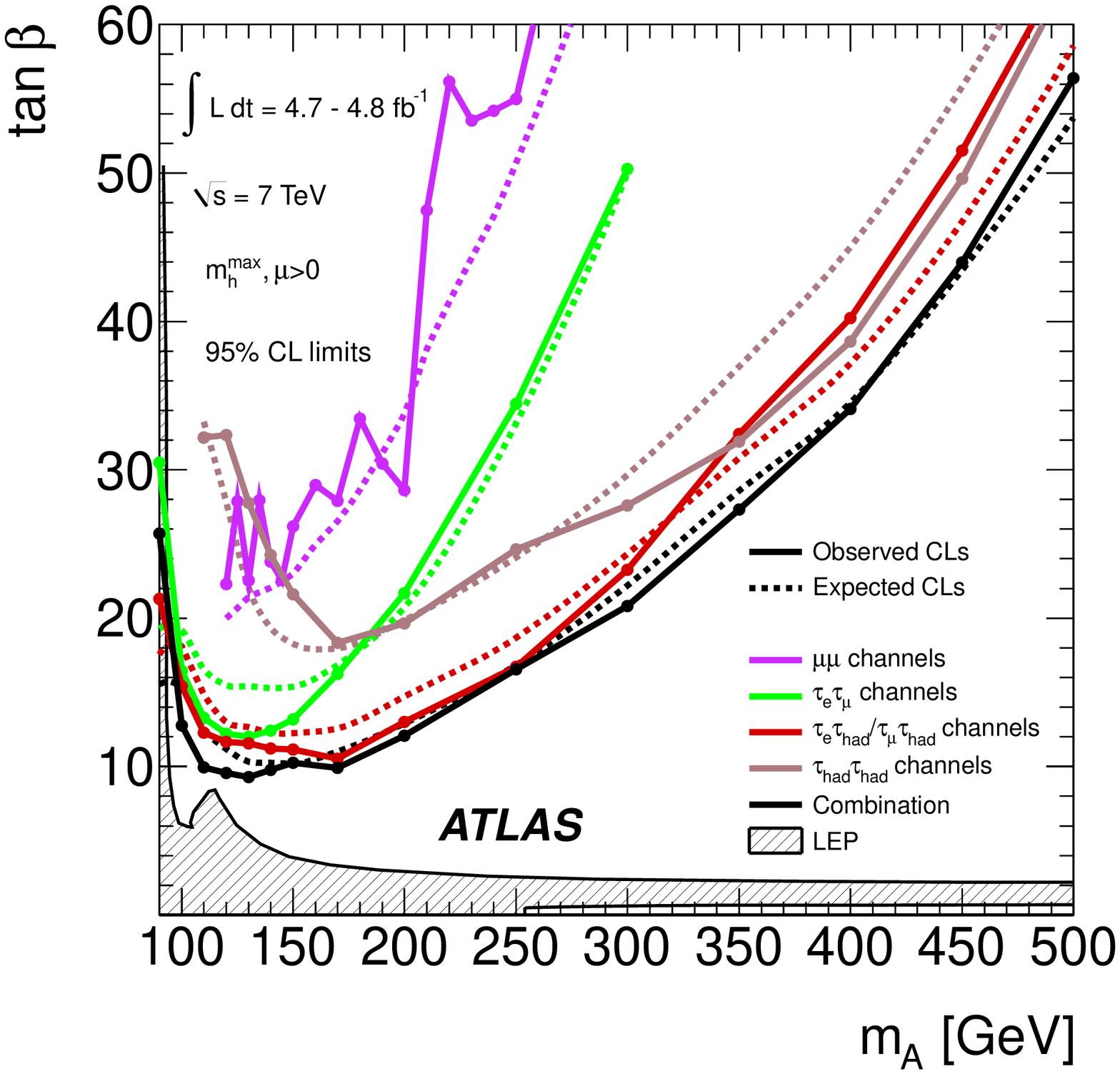}
  \caption{Expected (dashed line) and observed (solid line) \percent{95}
    CL  limits on \tanb{} as a function of $\MA$ for the statistical combination of all channels along with the $\pm 1\,\sigma$ and $\pm 2\,\sigma$ bands for the expected limit are shown on the left plot. Values of \tanb{} greater than the shown lines are excluded.
    The \percent{95}  CL limits  for the expected limit (dashed lines) and the observed limit (continuous lines) for each of the 
    $\mu\mu$, $\telec\tmuon$, $\tlep\thad$ and $\thad\thad$ channels and their statistical combination are shown on the right plot. The \percent{95} CL exclusion region from neutral MSSM Higgs boson searches performed at LEP\cite{LEPLimits} is shown in a hatched style.}
  \label{fig:tblimits}
\end{figure}

No significant excess of events above the background-only expectation is observed in the considered channels 
in \currentlumirange{} of $\sqrt{s}=\unit[7]{\TeV}$
proton--proton collision data. A \percent{95} CL upper limit on \tanb{} is set for each \MA{} point
using the frequentist method described in \secref{sec:statistics}. This is 
done using Higgs boson cross-sections calculated in the \mhmax{} scenario with $\mu > 0$~\cite{MSSMmhmax}.
Results for each of the $\mu\mu$, $\telec\tmuon$, $\tlep\thad$ and $\thad\thad$ final states, as well as for their statistical combination, can be seen in 
\figref{fig:tblimits}. The tightest constraint is at $\MA = \unit[130]{\GEVcc}$, where values of $\tanb > 9.3$ are excluded. The expected exclusion for the
same point is $\tanb > 10.3$. The exclusion of parameter space is significantly increased in comparison to earlier results by the ATLAS Collaboration~\cite{ATLASLimit} and complementary to the excluded parameter space from searches at LEP~\cite{LEPLimits}.
A significant portion of the MSSM parameter space that is not excluded
is still compatible with the assumption that the newly discovered particle
at the LHC is one of the neutral CP-even MSSM Higgs bosons \cite{theo125,theo125_v2}.
The lowest local $p_{0}$-values per channel are 0.014 ($2.2 \sigma$) at $\MA{}=\unit[125]{\GEVcc}$ for the $\mu\mu$ channel,
0.014 ($2.2 \sigma$) at $\MA{}=\unit[90]{\GEVcc}$ for the $\telec\tmuon$ channel, 0.067 ($1.5 \sigma$) at $\MA{}=\unit[90]{\GEVcc}$ for the $\tlep\thad$ channel
and 0.097 ($1.3 \sigma$) at $\MA{}=\unit[140]{\GEVcc}$ for the $\thad\thad$ channel.
The lowest local $p_{0}$ for the statistical combination of all channels is 0.004 ($2.7 \sigma$)
at $\MA=\unit[90]{\GEVcc}$. The significance of this excess is below $2 \sigma$, after considering the
look-elsewhere effect in the range $\unit[90]{\GEVcc} \leq \MA{} \leq \unit[500]{\GEVcc}$ and $5 \leq \tanb{} \leq 60$ \cite{LEE2D}.

\begin{figure}
  \centering
  \includegraphics[width=.45\textwidth]{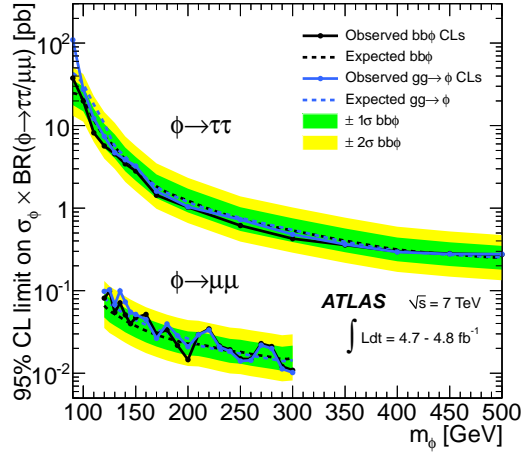}
  \caption{Expected (dashed line) and observed (solid line) \percent{95}
    CL limits on the cross-section for gluon-fusion and
    $b$-associated Higgs boson production times the branching ratio into $\tau$ and $\mu$ pairs,
respectively, along with the $\pm 1\,\sigma$ and $\pm 2\,\sigma$ bands for the
expected limit. The combinations of all $\tau\tau$ and $\mu\mu$ channels are shown. The difference
in the exclusion limits obtained for the gluon-fusion and the $b$-associated production modes is due
to the different sensitivity from the $b$-tagged samples.}
  \label{fig:xslimits}
\end{figure}

The outcome of the search is further interpreted in the 
generic case of a single scalar boson $\phi$ produced in either the 
gluon-fusion or $b$-associated production mode and decaying to \mumu{} or \tautau.
\Figref{fig:xslimits} shows \percent{95} CL limits based on this interpretation. 
The exclusion limits for the production cross-section times the branching ratio for a Higgs boson
decaying to \mumu{} or \tautau{} are shown as a function of the Higgs boson mass.

%% file: conclusions.tex
\section{Summary} \label{sec:conclusions}

A search is presented for the neutral Higgs bosons 
of the Minimal Supersymmetric Standard Model 
in proton--proton collisions at a centre-of-mass energy 
of \unit[7]{\TeV} with the ATLAS experiment at the LHC.
A significant portion of the available MSSM parameter space is
consistent with the assumption that the newly discovered particle at the
LHC is one of the neutral CP-even MSSM Higgs bosons.
The study is based on a data sample that corresponds 
to an integrated luminosity of \currentlumirange. 
The decay modes of the Higgs bosons considered are 
\Htomumu, \Htotautauemu, \Htotautautolh{} and \Htotautautohhnospace . 
The analysis selection criteria exploit the two main production
mechanisms in the MSSM, the gluon-fusion and $b$-associated production modes,
by introducing categories for event samples with and without
an identified \bjet .
Since no excess of events over the expected background is observed in the considered channels, 
\percent{95} CL limits are set in the $\MA$--$\tanb$ plane, 
excluding a significant fraction of the MSSM parameter space.

%% file: acknowledgements.tex
% Acknowledgements for papers with collision data
% Version 5-Nov-2012
\section{Acknowledgments}

We thank CERN for the very successful operation of the LHC, as well as the
support staff from our institutions without whom ATLAS could not be
operated efficiently.

We acknowledge the support of ANPCyT, Argentina; YerPhI, Armenia; ARC,
Australia; BMWF and FWF, Austria; ANAS, Azerbaijan; SSTC, Belarus; CNPq and FAPESP,
Brazil; NSERC, NRC and CFI, Canada; CERN; CONICYT, Chile; CAS, MOST and NSFC,
China; COLCIENCIAS, Colombia; MSMT CR, MPO CR and VSC CR, Czech Republic;
DNRF, DNSRC and Lundbeck Foundation, Denmark; EPLANET, ERC and NSRF, European Union;
IN2P3-CNRS, CEA-DSM/IRFU, France; GNSF, Georgia; BMBF, DFG, HGF, MPG and AvH
Foundation, Germany; GSRT and NSRF, Greece; ISF, MINERVA, GIF, DIP and Benoziyo Center,
Israel; INFN, Italy; MEXT and JSPS, Japan; CNRST, Morocco; FOM and NWO,
Netherlands; BRF and RCN, Norway; MNiSW, Poland; GRICES and FCT, Portugal; MERYS
(MECTS), Romania; MES of Russia and ROSATOM, Russian Federation; JINR; MSTD,
Serbia; MSSR, Slovakia; ARRS and MVZT, Slovenia; DST/NRF, South Africa;
MICINN, Spain; SRC and Wallenberg Foundation, Sweden; SER, SNSF and Cantons of
Bern and Geneva, Switzerland; NSC, Taiwan; TAEK, Turkey; STFC, the Royal
Society and Leverhulme Trust, United Kingdom; DOE and NSF, United States of
America.

The crucial computing support from all WLCG partners is acknowledged
gratefully, in particular from CERN and the ATLAS Tier-1 facilities at
TRIUMF (Canada), NDGF (Denmark, Norway, Sweden), CC-IN2P3 (France),
KIT/GridKA (Germany), INFN-CNAF (Italy), NL-T1 (Netherlands), PIC (Spain),
ASGC (Taiwan), RAL (UK) and BNL (USA) and in the Tier-2 facilities
worldwide.

%% file: atlas_authlist.tex
% ATLAS Collaboration author list
% Data extracted on 30-Nov-2012 for paper reference HIGG-2012-11
%\documentclass[11pt]{article}
%\usepackage{a4wide}\begin{document}
\begin{flushleft}
{\Large The ATLAS Collaboration}

\bigskip

G.~Aad$^{\rm 48}$,
T.~Abajyan$^{\rm 21}$,
B.~Abbott$^{\rm 111}$,
J.~Abdallah$^{\rm 12}$,
S.~Abdel~Khalek$^{\rm 115}$,
A.A.~Abdelalim$^{\rm 49}$,
O.~Abdinov$^{\rm 11}$,
R.~Aben$^{\rm 105}$,
B.~Abi$^{\rm 112}$,
M.~Abolins$^{\rm 88}$
% ...
\bigskip
\\
% ...
$^{11}$ Institute of Physics, Azerbaijan Academy of Sciences, Baku, Azerbaijan\\
$^{12}$ Institut de F{\'\i}sica d'Altes Energies and Departament de F{\'\i}sica de la Universitat Aut{\`o}noma de Barcelona and ICREA, Barcelona, Spain\\
% ...
$^{21}$ Physikalisches Institut, University of Bonn, Bonn, Germany\\
% ...
$^{48}$ Fakult{\"a}t f{\"u}r Mathematik und Physik, Albert-Ludwigs-Universit{\"a}t, Freiburg, Germany\\
$^{49}$ Section de Physique, Universit{\'e} de Gen{\`e}ve, Geneva, Switzerland\\
% ...
$^{88}$ Department of Physics and Astronomy, Michigan State University, East Lansing MI, United States of America\\
% ...
$^{105}$ Nikhef National Institute for Subatomic Physics and University of Amsterdam, Amsterdam, Netherlands\\
% ...
$^{111}$ Homer L. Dodge Department of Physics and Astronomy, University of Oklahoma, Norman OK, United States of America\\
$^{112}$ Department of Physics, Oklahoma State University, Stillwater OK, United States of America\\
% ...
$^{115}$ LAL, Universit{\'e} Paris-Sud and CNRS/IN2P3, Orsay, France\\
% ...
\end{flushleft}

%\end{document}
% Created with ./xml2latex.py